\documentclass[11pt]{article}
\usepackage[T1]{fontenc}
\usepackage{mathtools}
\usepackage{amssymb}
\usepackage{slashed}
\usepackage{multirow}
\usepackage{fullpage}
\usepackage[bottom]{footmisc}
\usepackage{longtable}
\usepackage{mathrsfs}
\usepackage{comment}
\usepackage[colorlinks=true,linkcolor=blue,citecolor=magenta,linktocpage=true]{hyperref}
\usepackage[numbers,sort,compress]{natbib}
\usepackage{tocloft}
\usepackage{titlesec}
\titleformat*{\section}{\large\bfseries}

\usepackage{array}

\def\be#1\ee{\begin{align}#1\end{align}}
\def\bsub#1\esub{\begin{subequations}#1\end{subequations}}
\def\nn{\nonumber}
\def\q{\qquad}
\def\f{\frac}

\def\ip{\lrcorner\,}
\def\ipp{\hbox{$\ip\!\!\!\!\!\;\ip$}}

\def\sq{\sqrt{q}}

\def\tr{\mathrm{tr}}
\def\tf{\text{\tiny{TF}}}

\def\de_\omega{\mathrm{D}}
\def\de{\mathrm{d}}

\def\C{\mathcal{C}}
\def\D{\mathcal{D}}
\def\E{\mathcal{E}}
\def\F{\mathcal{F}}

\def\I{\mathcal{I}}
\def\J{\mathcal{J}}

\def\M{\mathcal{M}}
\def\N{\mathcal{N}}
\def\O{\mathcal{O}}
\def\P{\mathcal{P}}

\def\R{\mathcal{R}}

\def\m0{(m_0)}
\def\mb0{(\bar{m}_0)}
\numberwithin{equation}{section}
\def\ra{\rangle}
\def\la{\langle}
\def\pe{\phantom{\ =}}
\def\ethb{\overline{\eth}}
\def\hateq{\;\widehat{=}}
\usepackage{pict2e}
\makeatletter
\DeclareRobustCommand{\loplus}{\mathbin{\mathpalette\dog@lsemi+}}
\DeclareRobustCommand{\lotimes}{\mathbin{\mathpalette\dog@lsemi{\times}}}
\DeclareRobustCommand{\roplus}{\mathbin{\mathpalette\dog@rsemi+}}
\DeclareRobustCommand{\rotimes}{\mathbin{\mathpalette\dog@rsemi{\times}}}

\newcommand{\dog@rsemi}[2]{\dog@semi{#1}{#2}{-90,90}}
\newcommand{\dog@lsemi}[2]{\dog@semi{#1}{#2}{270,90}}
\newcommand{\dog@semi}[3]{%
  \begingroup
  \sbox\z@{$\m@th#1#2$}%
  \setlength{\unitlength}{\dimexpr\ht\z@+\dp\z@\relax}%
  \makebox[\wd\z@]{\raisebox{-\dp\z@}{%
    \begin{picture}(1,1)
    \linethickness{\variable@rule{#1}}
    \roundcap
    \put(0.5,0.5){\makebox(0,0){\raisebox{\dp\z@}{$\m@th#1#2$}}}
    \put(0.5,0.5){\arc[#3]{0.5}}
    \end{picture}%
  }}%
  \endgroup
}
\newcommand{\variable@rule}[1]{%
  \fontdimen8  
  \ifx#1\displaystyle\textfont3\else
    \ifx#1\textstyle\textfont3\else
      \ifx#1\scriptstyle\scriptfont3\else
        \scriptscriptfont3\relax
  \fi\fi\fi
}
\makeatother

\begin{document}

\title{\Large{\textbf{\sffamily Symmetries of the gravitational scattering in the absence of peeling}}}
\author{\sffamily Marc Geiller${}^1$, Alok Laddha${}^2$, Céline Zwikel${}^3$
\date{\small{\textit{
$^1$Univ Lyon, ENS de Lyon, CNRS, Laboratoire de Physique, F-69342 Lyon, France\\
$^2$Chennai Mathematical Institute, H1, SIPCOT IT Park, Siruseri, Kelambakkam 603103, India\\
$^3$Perimeter Institute for Theoretical Physics,\\ 31 Caroline Street North, Waterloo, Ontario, Canada N2L 2Y5\\}}}}

\maketitle

\begin{abstract}
The symmetries of the gravitational scattering are intimately tied to the symmetries which preserve asymptotic flatness at null infinity. In Penrose's definition of asymptotic flatness, a central role is played by the notion of asymptotic simplicity and the ensuing peeling behavior which dictates the decay rate of the Weyl tensor. However, there is now accumulating evidence that in a generic gravitational scattering the peeling property is broken, so that the spacetime is not asymptotically-flat in the usual sense. These obstructions to peeling can be traced back to the existence of \textit{universal} radiative low frequency observables called ``tails to the displacement memory''. As shown by Saha, Sahoo and Sen, these observables are uniquely fixed by the initial and final momenta of the scattering objects, and are independent of the details of the scattering. The universality of these tail modes is the statement of the classical logarithmic soft graviton theorem. Four-dimensional gravitation scattering therefore exhibits a rich infrared interplay between tail to the memory, loss of peeling, and universal logarithmic soft theorems.

In this paper we study the solution space and the asymptotic symmetries for logarithmically-asymptotically-flat spacetimes. These are defined by a polyhomogeneous expansion of the Bondi metric which gives rise to a loss of peeling, and represent the classical arena which can accommodate a generic gravitational scattering containing tails to the memory. We show that while the codimension-two generalized BMS charges are sensitive to the loss of peeling at $\I^+$, the flux is insensitive to the fate of peeling. Due to the tail to the memory, the soft superrotation flux contains a logarithmic divergence whose coefficient is the quantity which is conserved in the scattering by virtue of the logarithmic soft theorem. In our analysis we also exhibit new logarithmic evolution equations and flux-balance laws, whose presence suggests the existence of an infinite tower of subleading logarithmic soft graviton theorems.

\end{abstract}

\thispagestyle{empty}
\newpage
\setcounter{page}{1}
\tableofcontents

\newpage

\section{Introduction}

The asymptotic structure of spacetime with vanishing cosmological constant, and in particular the notion of asymptotic flatness, plays a central role in the analysis of gravitational radiation \cite{Bondi:1960jsa,Bondi:1962px,Bondi:1962rkt,Sachs:1961zz,Sachs:1962wk,Newman:1961qr}. Since the pioneering work of Penrose and Geroch, asymptotic flatness is best understood in terms of the conformal compactification of the spacetime \cite{Penrose:1962ij,Penrose:1964ge,Penrose:1965am,Geroch1977,Frauendiener:2000mk}, which enables in particular to characterize the codimension-one asymptotic boundary known as (future and past) null infinity $\I^\pm$. This is formalized in the notion of ``asymptotic simplicity'', which also prescribes the decay rates of the Weyl tensor as one approaches infinity along a radial direction at fixed retarded or advanced time. This behavior is known as the ``peeling'' property \cite{Penrose:1962ij,Penrose:1965am,Geroch1977,Sachs:1961zz}, and in the physical spacetime it translates into fall-offs of the Weyl scalars given by \eqref{Sachs peeling}.

Bondi, Metzner, Van der Burg and Sachs have famously shown that asymptotic flatness is preserved by an infinite-dimensional group of symmetries in which the Abelian ideal of spacetime translations in the Poincar\'e group is enhanced to so-called supertranslations  \cite{Bondi:1960jsa,Bondi:1962px,Bondi:1962rkt,Sachs:1961zz,Sachs:1962wk}. This infinite-dimensional group is known as the BMS group. Rather remarkably, in the past two decades there have been major developments proposing extensions of the BMS group \cite{Barnich:2009se,Barnich:2010eb,Barnich:2011mi,Barnich:2011ct,Barnich:2013axa,Campiglia:2014yka,Campiglia:2015yka,Compere:2018ylh,Freidel:2021fxf,Geiller:2022vto,Geiller:2024amx}. The reason for this surge of interest in symmetries of asymptotically-flat spacetimes is due to the fact that they appear to have an intricate relationship with the symmetries of the gravitational scattering \cite{Strominger:2013jfa,He:2014laa,Strominger:2014pwa}. This connection between asymptotic symmetries at null infinity and conservation laws of classical scattering have motivated a closer inspection of the assumptions that went into the definition of the BMS group, to see if those assumptions could be relaxed while still preserving the notion of asymptotic flatness. Assuming the peeling behavior, several enhancements of BMS group (in particular the so-called extended eBMS and generalized gBMS groups) have therefore been proposed, and their implications for the classical (and quantum) S-matrix \cite{White:2014qia,Cachazo:2014fwa,Zlotnikov:2014sva,Kalousios:2014uva,Conde:2016rom,Campiglia:2016efb,Campiglia:2016jdj,Banerjee:2021cly,Freidel:2021dfs} and for the memory effects \cite{Pasterski:2015tva,Mitman:2020pbt,Nichols:2017rqr} have been analyzed.

In four spacetime dimensions and in the presence of massive sources, the radiative gravitational field sourced by a matter stress tensor exhibits remarkable universal properties at early and late retarded times at $\I^\pm$. These properties are quantified in terms of the well-known displacement memory, as well as sub-leading tails to the memory whose universality was proven in a series of seminal papers by Sahoo and Sen \cite{Sahoo:2018lxl}, and by Saha, Sahoo and Sen \cite{Saha:2019tub}. It turns out that these tails directly influence the asymptotic structure and imply a failure of the peeling, as was already pointed out in \cite{1986mgm..conf..365D,2002nmgm.meet...44C,1985FoPh...15..605W}. There is by now substantial evidence for the need to relax Penrose's notion of asymptotic flatness and the resulting peeling property \cite{Friedrich:1983vx,PhysRevD.19.3483,PhysRevD.19.3495,doi:10.1098/rspa.1981.0101,Andersson:1993we,Valiente-Kroon:2002xys,Kroon:2004me,Kehrberger:2021uvf,Kehrberger:2021vhp,Kehrberger:2021azo,Kehrberger:2024clh,Kehrberger:2024aak,Gajic:2022pst,Kehrberger:2023btg,Bieri:2023cyn}. This can be done by working in the context of so-called polyhomogeneous expansions of the Bondi metric, or logarithmically-asymptotically-flat spacetimes (LAF hereafter) \cite{1985FoPh...15..605W,Chrusciel:1993hx,Kroon:1998tu,ValienteKroon:2002gb,Godazgar:2020peu,Freidel:2024tpl}.

Since the definition of BMS asymptotic symmetries and their enhancement relies crucially on asymptotic simplicity, one can wonder whether a qualitative change in the notion of asymptotic flatness will impact the group of asymptotic symmetries. Two basic questions need to be answered in this regard. First, what is the group of asymptotic symmetries at\footnote{Throughout this paper we will assume that the spacetime is asymptotically-flat at $\I^-$ as is the case with generic scattering data with e.g. no incoming radiation.} $\I^+$ for logarithmically-asymptotically-flat spacetimes as opposed to asymptotically-flat spacetimes? Second, given this group of symmetries, what are the corresponding charges and fluxes? In the present work we answer these two questions. First, we show that the codimension-two supertranslation charges are insensitive to the fate of peeling, while the superrotation charges depend on the extra mode sourcing the loss of peeling. Then we show that the supertranslation and superrotation fluxes, computed via covariant phase space methods, are insensitive to the fate of peeling. These results imply that the gravitational contribution to the gBMS fluxes, used in establishing the equivalence of the classical soft theorems with the conservation laws, remains valid regardless of the fate of peeling at $\I^+$.


This result enables to put on a firmer footing earlier work on the relationship between classical logarithmic soft theorems and conservation laws associated to asymptotic symmetries. In \cite{Choi:2024ygx,Choi-Laddha-Puhm-WIP}, the authors have derived the classical superrotation flux for a massive scalar field coupled to gravity in $D=4$ spacetime dimensions. The resulting conservation law was then shown to be equivalent to the classical logarithmic soft theorem. However, this derivation was done in perturbative gravity around a specific Minkowski background, in the harmonic gauge, and under the assumption that the peeling was satisfied at $\I^+$. However, since the tail to the memory causes the loss of peeling, one cannot simply assume the flux formula to be given by the standard gBMS expression. It is this missing point in the analysis that our work aims at clarifying, and the result is precisely to show why this assumption is indeed correct.

Although the present work is solely concerned with classical charges which then are used in formulating conservation laws, we should note that in \cite{Agrawal:2023zea} the authors have quantized the superrotation charge obtained by assuming peeling, and then analyzed the corresponding Ward identities and their relationship with the quantum logarithmic soft theorem \cite{Sahoo:2018lxl}. The gravitational contribution to the superrotation flux used in \cite{Agrawal:2023zea} matches with the flux formula derived here in the absence of peeling\footnote{Scattering amplitudes in gravity in $D = 4$ dimensions satisfy a hierarchy of universal factorisation theorems. For example, given a scattering amplitude with incoming massive momenta $p_1,\dots,p_n$ and outgoing massive momenta $p'_1,\dots,p'_m$, and an outgoing graviton with momentum $k$, the soft expansion of the scattering amplitude $\M_{n+m+1}\big(p_1,\dots,p_n\rightarrow p'_1,\dots,p'_m,k=\omega(1,\hat{k})\big)$ factorises as
\be
\f{\M_{n+1}}{\M_n}=\f{S^{(0)}}{\omega}+(\ln\omega)S^{\ln}+\O(\omega^0).
\ee
It was shown in \cite{Sahoo:2018lxl} that $S^{(0)}$ and $S^{\ln}$ are infrared finite. The term 
$S^{(0)}=S^{(0)}(p_1,\dots,p_n,p'_1,\dots,p'_m,\hat{k})$ is the Weinberg soft factor and $S^{\ln}=S^{\ln}(p_1,\dots,p_n,p'_1,\dots,p'_m,\hat{k})$ is the quantum logarithmic soft factor. This quantum logarithmic soft factor $S^{\ln}$ \textit{differs} from the formula for the tail to the memory (i.e. the classical logarithmic soft factor) by additional terms, which arise from the fact that quantum amplitudes are computed using Feynman boundary conditions, while retarded boundary conditions are used in classical scattering.}. It is interesting to note that the charges which we obtain here in the absence of peeling also match the results obtained from a BRST analysis in \cite{Baulieu:2024oql}.

This paper is organized as follows. In sections \ref{subsec:tttm} and \ref{subsec:peeling} we review the relationship between late time modes of the gravitational radiation in $D=4$ spacetime dimensions and the peeling property of the Weyl scalars at $\I^+$. We try to summarize our understanding regarding loss of peeling and to contextualize this in light of the classical logarithmic soft theorems of Saha, Sahoo and Sen. In section \ref{lafspacetimes} we introduce a family of logarithmically-asymptotically-flat (LAF) spacetimes compatible with the loss of peeling expected from the classical logarithmic soft theorems. We work out the solution space, the evolution equations, the Newman--Penrose formulation, and identity the loss of peeling in the Weyl scalars. We also show that the Weyl--BMS (BMSW) group preserves the LAF solution space. In section \ref{sec:charges and all} we then compute the charges as well as the fluxes of all the gBMS generators at $\I^+$ when peeling is violated. In particular, we carefully analyze the soft superrotation flux and regularize its divergent contributions arising from the constant shear mode and the tail to the memory. We also give a corrected formula \eqref{new spin memory} for the spin memory which takes into account the constant shear mode (and an ambiguous regulator-dependent term) in the finite part of the soft superrotation flux. The appendices contain details on the asymptotic expansion for the spacetime metric and the Newman--Penrose formalism for LAF spacetimes. We also discuss there the issue of symplectic regularization of the divergencies in the charges. For the sake of completeness, we also present in appendix \ref{Christo argument} the classical argument by Christodoulou connecting the quadrupolar radiation due to massive sources in the past to the loss of peeling at $\I^+$.

\subsection{Tail to the memory}
\label{subsec:tttm}

One of the famous results in the theory of gravitational radiation is the discovery of the displacement memory effect \cite{1974SvA....18...17Z,Braginsky:1986ia,1987Natur.327..123B,Ludvigsen:1989kg,Christodoulou:1991cr,Blanchet:1992br,Thorne:1992sdb}. In any classical gravitational scattering involving $m$ incoming objects with momenta $p'_1,\dots,p'_m$ and $n$ outgoing objects with momenta $p_1,\dots,p_n$, the late time gravitational radiation is dominated by a static mode which is completely fixed (up to an overall phase which can be removed by changing the reference time at which the detector is switched on) by the incoming and outgoing momenta of the scattered particles\footnote{We allow the momenta to be massless, so that the displacement memory includes the null memory (also known as Christoudolou's non-linear memory) \cite{Bieri:2013ada}.}. The displacement memory formula can be obtained as follows. First, consider coordinates $x^\mu=(t,\vec{x})$ and the retarded time\footnote{The correction, which is typically in $\ln r$, is due to drag terms arising from the scattered objects. See section \ref{subsec:symplectic} for a more detailed discussion of their role.} $u=t-r-\O(G)$ where $r\coloneqq|\vec{x}|$. The radiative metric $e_{\mu\nu}$ is defined as
\be
e_{\mu\nu}(x)\coloneqq h_{\mu\nu}-\f{1}{2}\eta_{\mu\nu}\tr(h),
\q\q
h_{\mu\nu}\coloneqq\f{1}{2}\big(g_{\mu\nu}-\eta_{\mu\nu}\big).
\ee
Then, the displacement memory is the ``zero frequency mode'' of this radiative field, which can be obtained as\footnote{We have here momentarily reintroduced the speed of light $c$.}
\be\label{displacement memory}
e^0_{\mu\nu}(x^a)\coloneqq\lim_{r\to\infty}\,\lim_{u\to\infty}\,re_{\mu\nu}(x)=\f{2G}{c^3}\left(\sum_{i=1}^n\f{p^i_\mu p^i_\nu}{p_i\cdot n}-\sum_{j=1}^m\f{p'^j_\mu p'^j_\nu}{p'_j\cdot n}\right),
\ee
where $n^\mu=(1,\vec{x}/r)$ is the null vector that points to the celestial sphere, and where $x^a$ denote the angular coordinates. Schematically, the displacement memory is the radiative field which ``survives'' at null infinity as $\vert u\vert\to\infty$. This radiative observable owes its universality to the fact that supertranslations are a symmetry of the classical gravitational scattering in four spacetime dimensions\footnote{Universality here refers to the classical soft theorem, i.e. the displacement memory is independent of the details of the scattering and is completely determined by the scattering data.}.

More generally, if we now analyze the radiative field at late retarded time, i.e. for $u$ large and positive, we find when $D=4$ the asymptotic behavior
\be\label{u expansion of e}
\lim_{r\to\infty}re_{\mu\nu}(x)=e^0_{\mu\nu}(x^a)+\f{1}{u}e^1_{\mu\nu}(x^a)+\O\left(\f{\ln u}{u^{2}}\right).
\ee
Although the presence of a $1/u$ mode in the radiative field has been known for a long time \cite{1986mgm..conf..365D}, it is only recently, in a seminal series of articles by Saha, Sahoo and Sen \cite{Sahoo:2018lxl,Saha:2019tub}, that the term $e^1_{\mu\nu}$ was shown to be universal, exactly as the displacement memory! More precisely, building up on the factorisation theorem for the gravitational S-matrix given in \cite{Sahoo:2018lxl} and on the observations made in \cite{Laddha:2018vbn}, Saha, Sahoo and Sen proved in \cite{Saha:2019tub} that $e^1_{\mu\nu}$  is also given by a universal formula which is completely determined by the asymptotic momenta and masses of the scattering objects. This result is now known as the classical logarithmic soft graviton theorem, and can be written as
\be
\lim_{r\to\infty}\lim_{u\to\pm\infty}\,\mp\,ru^{2}\partial_ue_{\mu\nu}(x)\eqqcolon e^{1\pm}_{\mu\nu}(x^a).
\ee
The analytic expressions for $e^{1\pm}_{\mu\nu}(x^a)$ can be found e.g. in \cite{Saha:2019tub}. Without writing down the explicit formula, let us now emphasize the key aspects of this result.
\begin{enumerate}
\item The expressions for $e^{1\pm}_{\mu\nu}$ are uniquely determined by the set of incoming and outgoing momenta, and in particular do not depend on the details of the scattering nor on the spin of the scattered particles. This property is the same as for the displacement memory \eqref{displacement memory}.
\item The expression for $e^{1+}_{\mu\nu}$ depends on both the incoming and outgoing momenta involved in the scattering (see equation (1.7) in \cite{Sahoo:2021ctw}).
\item The expression for $e^{1-}_{\mu\nu}$ depends only on the incoming momenta (see equation (1.8) in \cite{Sahoo:2021ctw}).
\end{enumerate}
The reason behind this universality of the formula for $e^{1\pm}_{\mu\nu}$ lies in the universality of the long range interaction between the scattering objects (which also causes the orbital angular momentum of individual objects to diverge logarithmically at time-like infinity) and in the Coulombic drag that the outgoing radiation experiences due to the spacetime curvature effects. Rather remarkably, it was conjectured that among the subleading contributions in \eqref{u expansion of e} there are terms of the form $(\ln u)^m/u^{m+1}$ which are also universal and uniquely determined by the asymptotic momenta. A proof of this conjecture was given by Sahoo and Sen for $m=1$ in \cite{Sahoo:2021ctw}.

Let us now comment on the terminology, which refers to $e^1_{\mu\nu}$ as ``tail to the memory'' \cite{Laddha:2018vbn}. In the post-Newtonian literature, when radiation of binary systems is analyzed, the tail refers solely to the drag which ``slows down'' the outgoing radiation and causes a phase shift in the radiative field \cite{Blanchet:1993ec,Blanchet:2020ngx,Blanchet:2023pce,Trestini:2023wwg}. On the other hand, the tail to the memory as defined in \cite{Laddha:2018vbn} is a sum of this ``drag contribution'' and the contribution due to the logarithmic divergence of angular momenta of the scattering bodies. It is this sum which generates the $1/u$ mode as $u\to+\infty$, but only the latter contribution which generates the $1/u$ mode as $u\to-\infty$ (since the drag term is actually vanishing at $\I^+_-$) \cite{Sahoo:2021ctw}.

The next important property deriving from the result \eqref{u expansion of e} and playing a crucial role in the present work is that the presence of the $1/u$ tail to the memory implies a failure of the peeling behavior in such radiative spacetimes. To the best of our knowledge, the direct connection between the tail and the loss of peeling can be intuitively understood with an argument originally given by Christodoulou in \cite{2002nmgm.meet...44C}, in a form which we review in appendix \ref{Christo argument}. This was then refined and established rigorously by Kehrberger in a remarkable series of articles \cite{Kehrberger:2021uvf,Kehrberger:2021vhp,Kehrberger:2021azo,Kehrberger:2024clh,Kehrberger:2024aak} (see also \cite{Gajic:2022pst,Kehrberger:2023btg}). Let us now explain the issues pertaining to the peeling property.

\subsection{Loss of peeling}
\label{subsec:peeling}

Building up on the foundational work introducing the framework of asymptotic flatness at null infinity \cite{Bondi:1960jsa,Bondi:1962px,Bondi:1962rkt,Sachs:1961zz,Sachs:1962wk,Newman:1961qr}, Penrose has proposed in \cite{Penrose:1962ij} the notion of ``asymptotic simplicity'', characterized by the requirement of smoothness of the conformal structure across null infinity. It was then realized in \cite{Penrose:1965am} (see also \cite{Geroch1977}) that the smoothness of the conformal fields is intimately related to the vanishing at $\I^\pm$ of the Weyl tensor of the conformal spacetime metric. In terms of the physical fields, this translates into specific fall-off rates for the Weyl tensor. These rates, known as the ``Sachs peeling'' property, were furthermore conjectured to be characteristic of the asymptotic radiation emitted by isolated self-gravitating systems \cite{Sachs:1961zz,Kroon:2016ink}. Explicitly, denoting the Weyl scalars by $\Psi_k$ for $k\in\{0,\dots,4\}$, the peeling property is characterized by the asymptotic behavior
\be\label{Sachs peeling}
\Psi_k\big|_{r\to\infty}=\Psi_k^0\,r^{k-5}+\O\big(r^{k-6}\big),
\ee
which is here understood at $\I^+$.

Asymptotically-flat spacetimes satisfying the peeling property are the arena in which asymptotic symmetries and the BMS group were discovered and extensively studied \cite{Bondi:1960jsa,Bondi:1962px,Bondi:1962rkt,Sachs:1962zza}. There results were then generalized to allow for more general boundary conditions, and there is by now substantial evidence that the so-called generalized BMS group (gBMS hereafter) \cite{Banks:2003vp,Barnich:2009se,Barnich:2010eb,Barnich:2011mi,Barnich:2011ct,Barnich:2013axa,Campiglia:2014yka,Campiglia:2015yka,Compere:2018ylh,Freidel:2021fxf}, which is the semi-direct product of supertranslations with smooth diffeomorphisms of celestial sphere (or superrotations), is a symmetry of the classical gravitational scattering\footnote{One can also consider simply the so-called extended BMS group eBMS, which is the semi-direct product of supertranslations with the loop group generated by meromorphic vector fields.}. In particular, the conservation law associated with the superrotations and the corresponding angular momentum charges produces the formula for the so-called spin memory \cite{Pasterski:2015tva}.

However, since the gBMS symmetries of the scattering have only been studied in the case of spacetimes satisfying the peeling property, it is natural to ask what happens when this assumption is dropped. This is a particularly important question given the fact that the fate of peeling remains to this day a contentious point (see \cite{Friedrich:2017cjg} for a detailed account of the issues and relevant references). More precisely, the question is whether generic radiative spacetimes (e.g. arising from the scattering of interacting massive sources in the far past) satisfy peeling in the future. There are indeed strong indications that this is not the case \cite{Friedrich:1983vx,PhysRevD.19.3483,1985FoPh...15..605W,PhysRevD.19.3495,1986mgm..conf..365D,doi:10.1098/rspa.1981.0101,2002nmgm.meet...44C,Andersson:1993we} (more recently, see also \cite{Valiente-Kroon:2002xys,Kroon:2004me,Kehrberger:2021uvf,Kehrberger:2021vhp,Kehrberger:2021azo,Kehrberger:2024clh,Kehrberger:2024aak,Gajic:2022pst,Kehrberger:2023btg,Bieri:2023cyn}). In the typical study of the gravitational radiation produced by binary systems, peeling is recovered as a consequence of the assumption of past-stationarity of the metric \cite{Blanchet:1985sp,Blanchet:1986dk}. However, this is precisely the assumption which fails in the presence of interacting and unbound particles in the far past (e.g. for hyperbolic encounters \cite{PhysRevD.1.1559,PhysRevD.19.3483,PhysRevD.19.3495,1978ApJ...224...62K,Hait:2022ukn,Bini:2023gaj}). Since the results of Saha, Sahoo and Sen \cite{Sahoo:2018lxl,Saha:2019tub} describe a completely general scattering situation, it should not come as a surprise that they also imply a failure of peeling. More precisely, this can be traced back to the presence of the $1/u$ tail to the memory, as already argued by Christodoulou \cite{2002nmgm.meet...44C}. We also note that Damour had reached the same conclusion by working in a multipolar post-Minkowskian expansion in harmonic gauge \cite{1986mgm..conf..365D}.


In order to make things concrete, consider a state of initial massive particles with masses $m_1,\dots,m_n$ and momenta $p_1,\dots,p_n$, and assume no incoming radiation at $\I^-$. This condition implies that $\I^-$ is smooth, with the usual $1/r$ expansion of the radiative metric and peeling of the Weyl tensor at $\I^-$. In $D=4$ spacetime dimensions, the asymptotic trajectory of the massive particles as $t\to-\infty$ can be parametrized as
\be\label{loglate}
x_i^\mu(t)=b_i^\mu+v_i^\mu t+c_i^\mu\ln|t|,
\ee
where $b_i^\mu$ and $v_i^\mu$ specify the initial configuration of the $i$-th particle, and $c_i^\mu$ is the deviation of the particle's asymptotic trajectory from the free trajectory due to the gravitational interaction with all the other particles. Note that the total relativistic angular momentum $M^{\mu\nu}=\sum_{i}p^{[\mu}_ix^{\nu]}_i$ diverges due to the presence of this logarithmic term. Using \eqref{loglate}, which implies absence of stationarity in the far past, and assuming smoothness of $\I^-$ with no incoming radiation, Christodoulou argued that the regularity of the metric at $i^0$ implies that
\begin{enumerate}
\item The asymptotic fall-off of the shear $C_{ab}(u,x^a)$ as $u\to-\infty$ is $\O(u^{-1})$.
\item The presence of this $1/u$ mode in the shear implies the loss of peeling at $\I^+$.
\end{enumerate}
The precise way in which the loss of peeling manifests itself is through modified fall-off rates for $\Psi_0$ and $\Psi_1$, which instead of \eqref{Sachs peeling} are given respectively by \eqref{Psi0} and \eqref{Psi1}. The detailed steps of the argument of Christodoulou are summarized in appendix \ref{Christo argument}.

The present work aims at exploring asymptotic symmetries in the ``worst case scenario'', i.e. after accepting the results of Damour \cite{1986mgm..conf..365D}, Christodoulou \cite{2002nmgm.meet...44C} and Kehrberger \cite{Kehrberger:2023btg} (among others), which assert that radiative spacetimes obtained from the scattering of massive particles do not satisfy peeling. This can be done by starting from a class of radiative spacetimes in Bondi gauge which break peeling in a controlled manner compatible with the above-mentioned results. In such spacetimes, the radial expansion near null infinity is called ``polyhomogeneous'' since it involves terms in $\ln r$ \cite{1985FoPh...15..605W,Chrusciel:1993hx,Kroon:1998tu,ValienteKroon:2002gb} (a polyhomogeneous expansion is also required in three-dimensional Einstein--Maxwell theory \cite{Barnich:2015jua,Bosma:2023sxn}). More precisely, we will focus on polyhomogeneous spacetimes which are asymptotically-flat and where the logarithmic branches only impact the fall-offs of $\Psi_0$ and $\Psi_1$, and we will refer to them as logarithmically-asymptotically-flat (LAF) spacetimes.

In these LAF spacetimes, we want to understand how the radiative phase space is modified by the loss of peeling and what are the charges associated to the generators of gBMS symmetries. Clearly, the asymptotic symmetries must contain the BMS supertranslations since the displacement memory remains a well-defined observable even in the absence of peeling. Since we have argued above that the tail to the memory is another universal radiative observable, it is natural to ask how it interplays with the gBMS asymptotic symmetries. However, since the tail to the memory induces a failure of peeling, this question must necessarily be phrased in the context of LAF spacetimes. 

As our starting point, we consider LAF spacetimes which break peeling in a way compatible with the results of Christodoulou \cite{2002nmgm.meet...44C} and Kehrberger \cite{Kehrberger:2021uvf,Kehrberger:2021vhp,Kehrberger:2021azo,Kehrberger:2024clh,Kehrberger:2024aak}. The polyhomogeneous ansatz for these spacetimes in Bondi gauge is described in detail in section \ref{sec:solution space}. Then we show that the gBMS charges are sensitive to the fate of peeling, while on the other hand the soft superrotation flux is agnostic to the loss of peeling in spite of containing the tail contribution responsible for the logarithmic soft graviton theorem \cite{Agrawal:2023zea,Choi:2024ygx,Choi-Laddha-Puhm-WIP}. In summary, this result shows that gBMS\footnote{For the sake of generality, we are actually going to study the so-called BMS--Weyl (BMSW) group, which in addition to gBMS also contains the freedom to perform conformal transformations of the codimension-2 boundary volume $\sq$.} is a symmetry of the classical scattering even in the absence of peeling. Along the way, we also study the generic form of the new logarithmic flux-balance laws which appear in the LAF spacetimes. This sets the stage for a future study of the tower of subleading logarithmic soft theorems.

Interestingly, the hamiltonian viewpoint on asymptotic symmetries also reveals that the presence of BMS symmetries is insensitive to the fate of peeling. This has been investigated around spatial infinity in \cite{Fuentealba:2023syb} (see footnote 7) and \cite{Henneaux:2018hdj,Henneaux:2018mgn,Henneaux:2019yqq}.

\section{Logarithmically-asymptotically-flat spacetimes}
\label{lafspacetimes}

In this section we study a class of logarithmically-asymptotically-flat (LAF) spacetimes near null infinity in the Bondi gauge. For this, we first construct the solution space using a polyhomogeneous expansion, and then derive the Einstein evolution equations along $\I^+$. We then compute the Weyl scalars to identify the loss of peeling behavior. Finally, we analyze the BMSW asymptotic symmetries and their action on the LAF solution space. These ingredients are then used in section \ref{sec:charges and all} in order to compute the symplectic structure, the charges and the fluxes.

\subsection{Solution space and flux-balance laws}
\label{sec:solution space}

We start with the line element in Bondi gauge
\be\label{Bondi gauge}
\de s^2=e^{2B}\f{V}{r}\de u^2-2e^{2B}\de u\,\de r+\gamma_{ab}(\de x^a-U^a\de u)(\de x^b-U^b\de u).
\ee
For the angular metric we then choose a polyhomogeneous expansion of the form
\be\label{angular metric}
\gamma_{ab}=r^2q_{ab}\sqrt{1+\f{[\C\C]}{2r^2}}+r\C_{ab},
\q\q
r\C_{ab}=rC_{ab}+D_{ab}+\sum_{n=1}^\infty\sum_{m=0}^{n+1}\f{E^{n,m}_{ab}(\ln r)^m}{r^n},
\ee
where all the tensors appearing in $\C_{ab}$ are trace-free in $q_{ab}$ and we denote $[\C\C]=\C^{ab}\C_{ab}$. This expansion solves the Bondi--Sachs determinant condition $\sqrt{\gamma}=r^2\sq$, where $\gamma\coloneqq\det(\gamma_{ab})$ and similarly for $q$. In appendix \ref{app:angular expansion} we give the explicit expansion of this angular metric up to $n=3$, making in particular all the logarithmic terms explicit. It is important to stress at this point that we have chosen the upper bound on the admissible logarithmic terms in \eqref{angular metric} to be $m_\text{max}=n+1$ at every order $r^{-n}$. This choice, which defines in turn our class of LAF spacetimes, is motivated by the work \cite{Kehrberger:2024aak} (see also \cite{Kroon:1998dv}), which shows that for infalling masses on hyperbolic orbits near $i^-$ a non-vanishing term $E^{1,2}_{ab}$ is produced\footnote{The work \cite{Chrusciel:1993hx} (see also section 6.10 of \cite{Chrusciel:2002cp}) shows that an even more general polyhomogeneous expansion leads to a well-defined solution space in Bondi gauge. This corresponds to an angular metric of the form
$$\gamma_{ab}=r^2q_{ab}+r\Big\{C_{ab}+C^1_{ab}(\ln r)\Big\}+r^0\left\{D_{ab}+\f{1}{4}q_{ab}[CC]+D^1_{ab}(\ln r)+D^2_{ab}(\ln r)^2\right\}+\O(r^{-1}).$$
Logarithmic terms at the same order as the shear were also considered in \cite{Capone:2021ouo}. When comparing with \eqref{angular metric}, the time evolution of the new logarithmic terms is simply $\partial_uC^1_{ab}=\partial_uD^1_{ab}=\partial_uD^2_{ab}=0$. The evolution \eqref{Dab EOM} of $D_{ab}$, however, now gets replaced by $4\partial_uD_{ab}+2D_{\la a}D^cC^1_{b\ra c}-RC^1_{ab}=0$. In this enlarge polyhomogeneous solution space there are no additional asymptotic symmetries in Bondi gauge (see however \cite{Geiller:2022vto,Geiller:2024amx} for potentially new symmetries in the partial Bondi gauge). More importantly, there is so far no physical motivation for studying such an enlarged class of LAF spacetimes, which is why we stick to \eqref{angular metric} in the present work.}. Moreover we also argue in appendix \ref{Christo argument}, along the lines of the argument made by Christodoulou in \cite{2002nmgm.meet...44C}, that if the $u$ behavior of the shear is $C_{ab}=\#u^{-1}+\#(\ln u)u^{-2}+\O(u^{-3})$, as indeed implied by the classical soft graviton theorem \cite{Sahoo:2018lxl,Saha:2019tub}, then the Weyl scalar $\Psi_0$ must contain terms in $r^{-5}(\ln r)^2$. We will show in equation \eqref{Psi0} of section \ref{Weyl and NP} that such terms in $\Psi_0$ are indeed produced by $E^{1,2}_{ab}$.

Starting with the line element \eqref{Bondi gauge} and the polyhomogeneous ansatz \eqref{angular metric}, we can now solve the Einstein equations in a $1/r$ expansion. Note that here and in the rest of this work it is implicitly understood that $1/r$ expansions also contain terms of the form $(\ln r)^m$. Assuming the boundary conditions $B=\O(r^{-2})$, $U^a=\O(r^{-2})$, and $\partial_uq_{ab}=0$, the four hypersurface Einstein equations $G_{rr}=G_{ra}=G_{ru}=0$ are solved by the expansions
\bsub\label{expansions of BUV}
\be
B&=-\f{1}{32r^2}[CC]-\f{1}{12r^3}[CD]+\sum_{n=4}^\infty\sum_{m=0}^{n-2}\f{B_{n,m}(\ln r)^m}{r^n},\\
U^a
&=-\f{1}{2r^2}D_bC^{ab}+\f{1}{r^3}\big(N^a+U^a_{3,1}(\ln r)\big)+\sum_{n=4}^\infty\sum_{m=0}^{n-2}\f{U^a_{n,m}(\ln r)^m}{r^n},\\
V&=-\f{R}{2}r+2M+\sum_{n=1}^\infty\sum_{m=0}^n\f{V_{n,m}(\ln r)^m}{r^n}.
\ee
\esub
Here $R$ denotes the Ricci scalar of the metric $q_{ab}$, and the integration constants $M(u,x)$ and $N^a(u,x)$ represent as usual the mass and angular momentum aspects. The first terms in the on-shell expansions are gathered in appendix \ref{app:metric expansion}. It is interesting to note that even in the absence of logarithmic terms in the angular metric \eqref{angular metric} (which we have introduced as a choice), the full spacetime metric still develops logarithmic terms because $U^a_{3,1}\neq0$ when $D_{ab}\neq0$. These are the logarithmic terms which result in the loss of peeling in $\Psi_0$ and $\Psi_1$.

Having solved the hypersurface equations with the expansion \eqref{expansions of BUV}, we are now left with evolution equations. For the mass and angular momentum aspects, the latter are contained in the Einstein equations
\bsub
\be
G_{uu}\big|_{\O(r^{-2})}&=-2\partial_u\M+D_a\J^a+\f{1}{2}C_{ab}\N^{ab},\\
G_{ua}\big|_{\O(r^{-2})}&=-\big(D^b\partial_uD_{ab}\big)(\ln r)-\partial_u\P_a+\partial_a\M+\widetilde{\partial}_a\widetilde{\M}+C_{ab}\J^b-\f{1}{2}D^b\partial_uD_{ab},\label{Gua EOM}
\ee
\esub
where $(\M,\P_a,\J_a,\N_{ab})$ are the quantities defined in \eqref{covariant functionals} via the Weyl scalars. Using the fact that $\partial_uD_{ab}=0$, as shown below in \eqref{EOM D and Enn1}, this can be rewritten more compactly as
\bsub
\be
\partial_u\M&=\f{1}{2}D_a\J^a+\f{1}{4}C_{ab}\N^{ab},\label{curly M EOM}\\
\partial_u\P_a&=\partial_a\M+\widetilde{\partial}_a\widetilde{\M}+C_{ab}\J^b.\label{curly P EOM}
\ee
\esub
We therefore see that the functional form of the flux-balance laws for the mass and the angular momentum is not modified by the presence of logarithmic terms, which only appear indirectly through the contribution of $D_{ab}$ in \eqref{curly P}. However, this does not change the evolution equation for the momentum $\P_a$ since $D_{ab}$ is conserved by virtue of \eqref{EOM D and Enn1}.

\subsection{Logarithmic flux-balance laws}
\label{sec:log fb laws}

We now study the remaining flux-balance laws, which are encoded in the trace-free angular Einstein equations
\be
G_{ab}^\tf\coloneqq G_{ab}-\f{1}{2}(\gamma^{cd}G_{cd})\gamma_{ab}=0.
\ee
Expanding these equations, we find that they are of the form
\be\label{angular EOMs compact}
G_{ab}^\tf\big|_{\O(r^{-1})}=-\partial_uD_{ab},
\q\q
G_{ab}^\tf\big|_{\O(r^{-{n+1}})}=\sum_{m=0}^{n+1}\mathrm{EOM}\big(E^{n,m}_{ab}\big)(\ln r)^m,
\ee
where $\mathrm{EOM}\big(E^{n,m}_{ab}\big)$ contains the evolution equation for $E^{n,m}_{ab}$. The explicit expressions for some of these evolution equations up to $n=3$ are given in appendix \ref{app:angular Einstein expansion}. In particular, one can deduce that the evolution equations $\partial_uE^{n,m}_{ab}$ exhibit a different pattern for $m=n+1$, $m=n$, and $m<n$.

At every order $n$ the highest logarithmic term, which corresponds to $m=n+1$, is always a constant of the motion, i.e. we have
\be\label{EOM D and Enn1}
\partial_uD_{ab}=0,
\q\q
\partial_uE^{n,n+1}_{ab}=0.
\ee
One can see from \eqref{angular EOMs compact} that $D_{ab}$ would be constant even in the presence of massless matter whose stress tensor falls off as $T_{ab}=\O(r^{-2})$. Similarly, the tensors $E^{n,n+1}_{ab}$ would be conserved even if the stress tensor had logarithmic contributions of the form $T_{ab}\big|_{\O(r^{-{n+1}})}=\O\big(r^{-n}(\ln r)^n\big)$. Then the second to highest logarithmic term, which corresponds to $m=n$, evolves for $n=1$ and $n\geq2$ as
\be\label{EOM E11 and Enn}
\partial_uE^{1,1}_{ab}=\f{1}{6}\big(\Delta-R\big)D_{ab},
\q
\partial_uE^{n,n}_{ab}=-\f{n}{2(n+2)(n-1)}\left(\Delta+\f{1}{2}(n^2+n-4)R\right)E^{n-1,n}_{ab},
\ee
where $E^{n-1,n}_{ab}$ is itself an absolutely conserved quantity by virtue of \eqref{EOM D and Enn1}. The evolution equations \eqref{EOM D and Enn1} and \eqref{EOM E11 and Enn} can therefore straightforwardly be integrated to obtain absolutely conserved quantities, i.e. quantities which are conserved even in the presence of radiation.

For $m=n-1$, we then see in \eqref{EOM 21} that $\partial_uE^{2,1}_{ab}$ is sourced by the shear, while \eqref{EOM E32} shows that $\partial_uE^{3,2}_{ab}$ is sourced by the news. It is then reasonable to conjecture that all the remaining equations $\partial_uE^{n,m<n}_{ab}$ for $n\geq3$ are sourced by the news as well. The upshot of this analysis is that the inclusion of logarithmic terms for $m\geq1$ in \eqref{angular metric} gives rise to a new infinite tower of logarithmic flux-balance laws. It will be interesting for future work to study which part of these flux-balance laws is universal and which part is theory-dependent. Given that the tower of flux-balance laws corresponding to $E^{n,0}_{ab}$ is related to the $(\text{sub})^{n+1}$-leading soft graviton theorems and to a $w_{1+\infty}$ algebra of subleading symmetries \cite{Guevara:2021abz,Strominger:2021lvk,Ball:2021tmb,Freidel:2021ytz,Adamo:2021lrv,Geiller:2024bgf}, it will also be important to investigate whether a similar algebraic structure arises from the logarithmic contributions in \eqref{angular metric}. We postpone this investigation to future work.

\subsection{Weyl scalars and absence of peeling}
\label{Weyl and NP}

We now study the Weyl scalars and their Newman--Penrose evolution equations. Our goal is to show how the presence of $D_{ab}$ leads to a loss of peeling in $\Psi_0$ and $\Psi_1$, and how the logarithmic terms $E^{n,m\neq0}_{ab}$ in \eqref{angular metric} enter $\Psi_0$.

Let us consider the outgoing vector $\ell$ tangent to the null geodesics intersecting $\I^+$ along 2-spheres at constant $u$, and the ingoing vector $n$ tangent to $\I^+$ and transverse to $\ell$. These are given explicitly by
\be\label{null vectors}
\ell^\mu\partial_\mu\coloneqq\partial_r,
\q\q
n^\mu\partial_\mu\coloneqq e^{-2\beta}\left(\partial_u+\f{V}{2r}\partial_r+U^a\partial_a\right)=\partial_u-\f{R}{4}\partial_r+\O(r^{-1}),
\ee
and are such that $\ell^\mu n_\mu=-1$. Let us then consider the tetrad $e_i$ of null vectors
\be\label{null frames}
e_1\coloneqq\ell,
\q\
e_2\coloneqq n,
\q\
e_3\coloneqq m=m^a\partial_a=\sqrt{\f{\gamma_{\theta\theta}}{2\gamma}}\left(\f{\sqrt{\gamma}+i\gamma_{\theta\phi}}{\gamma_{\theta\theta}}\,\partial_\theta-i\partial_\phi\right),
\q\
e_4\coloneqq\bar{m}.
\ee
These are such that $\ell^\mu n_\mu=-1=-m^\mu\bar{m}_\mu$ with all the other contractions vanishing. The metric is $g^{\mu\nu}=e^\mu_ie^\nu_j\eta^{ij}$, where $\eta^{12}=-1$ and $\eta^{34}=+1$, and the angular metric is $\gamma_{ab}=2m_{(a}\bar{m}_{b)}$. The radial expansion of $m^a$ is\footnote{As explained in \cite{Geiller:2024bgf}, it is always possible to perform Lorentz transformations in order to set the spin coefficient $\epsilon=0$, and therefore $\epsilon_2=0$. This has the advantage of simplifying the expansion of $m^a$, and therefore of removing some terms in the expansion of $\Psi_0$, as can be seen in \eqref{Psi0}. The disadvantage however is that in this case the tetrad cannot be written in closed form as in \eqref{null frames}. For the sake of simplicity we therefore stick to the Bondi tetrad \eqref{null frames} in this work. Note however that this is a gauge freedom with no physical consequences.}
\be
m^a=\f{m^a_1}{r}+\f{m^a_2}{r^2}+\O(r^{-3}),
\quad\,\,
m^a_1=\sqrt{\f{q_{\theta\theta}}{2q}}\left(\f{\sq+iq_{\theta\phi}}{q_{\theta\theta}}\,\delta^a_\theta-i\delta^a_\phi\right),
\quad\,\,
m^a_2=\bar{m}^a_1\sigma_2+2m^a_1\epsilon_2,
\ee
where $\sigma_2$ and $\epsilon_2$ are given in \eqref{alpha sigma epsilon}. Using this we can then compute and expand the Weyl scalars. We find
\bsub\label{Weyl scalars}
\be
\Psi_0&\coloneqq-W_{\ell m\ell m}=\f{1}{r^4}D_{ab}m^a_1m^b_1\label{Psi0}\\
&\q\q\q\ +\f{1}{r^5}\Big(\E^{1,0}_{ab}+4\epsilon_2D_{ab}+\big(3E^{1,1}_{ab}-5E^{1,2}_{ab}\big)(\ln r)+3E^{1,2}_{ab}(\ln r)^2\Big)m^a_1m^b_1+\O(r^{-6}),\quad\nn\\
\Psi_1&\coloneqq-W_{\ell n\ell m}=\f{1}{r^4}\Big(D^bD_{ab}(\ln r)+\mathcal P_a\Big)m^a_1+\O(r^{-5})=\f{1}{r^4}\Big(\Psi_1^{0,1}(\ln r)+\Psi_1^0\Big)+\O(r^{-5}),\label{Psi1}\\
\Psi_2&\coloneqq-W_{\ell m\bar{m}n}=-\f{1}{2}\big(W_{\ell n\ell n}-W_{\ell nm\bar{m}}\big)=\f{1}{r^3}\big(\M+i\widetilde{\M}\big)+\O(r^{-4})=\f{\Psi_2^0}{r^3}+\O(r^{-4}),\label{Psi2}\\
\Psi_3&\coloneqq-W_{n\bar{m}n\ell}=\f{1}{r^2}\J_a\bar{m}^a_1+\O(r^{-3})=\f{\Psi_3^0}{r^2}+\O(r^{-3}),\label{Psi3}\\
\Psi_4&\coloneqq-W_{n\bar{m}n\bar{m}}=\f{1}{r}\N_{ab}\bar{m}^a_1\bar{m}^b_1+\O(r^{-2})=\f{\Psi_4^0}{r^1}+\O(r^{-2}),\label{Psi4}
\ee
\esub
where the so-called covariant functionals are given by
\bsub\label{covariant functionals}
\be
\E^{1,0}_{ab}&\coloneqq3E^{1,0}_{ab}-\f{3}{16}[CC]C_{ab}-\f{5}{2}E^{1,1}_{ab}+E^{1,2}_{ab},\label{curly E}\\
\P_a&\coloneqq-\f{3}{2}N_a+\f{3}{32}\partial_a[CC]+\f{3}{4}C_{ab}D_cC^{bc}-\f{4}{3}D^bD_{ab},\label{curly P}\\
\M&\coloneqq M+\f{1}{16}\partial_u[CC],\label{curly M}\\
\widetilde{\M}&\coloneqq\f{1}{4}\left(D_aD_b-\f{1}{2}N_{ab}\right)\widetilde{C}^{ab},\label{curly M dual}\\
\J_a&\coloneqq\f{1}{2}D^bN_{ab}+\f{1}{4}\partial_aR,\label{curly J}\\
\N_{ab}&\coloneqq\f{1}{2}\partial_uN_{ab}.\label{curly N}
\ee
\esub
First, one can see how the new terms $D_{ab}$ and $E^{n,m}_{ab}$ in the metric \eqref{angular metric} enter \eqref{curly E} and \eqref{curly P}. Second, and most importantly, one can see how these terms also cause the failure of peeling in $\Psi_0$ and $\Psi_1$. In particular, even when $D_{ab}=0$ we see that $\Psi_0$ contains terms in $r^{-5}(\ln r)$ and $r^{-5}(\ln r)^2$ sourced by $E^{1,1}_{ab}$ and $E^{1,2}_{ab}$. From the equations of motion \eqref{EOM D and Enn1} and \eqref{EOM E11 and Enn}, we also know that the coefficients of these terms are respectively linear in $u$ and constant.

In the Newman--Penrose formalism the flux-balance laws follow from the Bianchi identities, and play a central role in Christodoulou's argument relating the tail term in the shear to the loss of peeling, as summarized in appendix \ref{Christo argument}. These equations are
\bsub
\be
n^\mu\partial_\mu\Psi_2&=m^\mu\partial_\mu\Psi_3+2(\tau-\beta)\Psi_3+3\mu\Psi_2-2\nu\Psi_1-\sigma\Psi_4,\label{NP2}\\
n^\mu\partial_\mu\Psi_1&=m^\mu\partial_\mu\Psi_2+3\tau\Psi_2+2(\mu-\gamma)\Psi_1-\nu\Psi_0-2\sigma\Psi_3,\label{NP1}\\
n^\mu\partial_\mu\Psi_0&=m^\mu\partial_\mu\Psi_1+2(2\tau+\beta)\Psi_1+(\mu-4\gamma)\Psi_0-3\sigma\Psi_2.\label{NP0}
\ee
\esub
Expanding \eqref{NP2} we find
\be\label{Psi20 EOM}
\partial_u\Psi_2^0=\eth\Psi_3^0-\sigma_2\Psi_4^0,
\ee
which is the same as the sum of \eqref{curly M EOM} and its dual. Expanding \eqref{NP1} we get
\be\label{Psi10 EOM}
\partial_u\Psi_1^{0,1}=0,
\q\q
\partial_u\Psi_1^0=\eth\Psi_2^0-2\sigma_2\Psi_3^0.
\ee
The first equation is the contraction of the $(\ln r)$ term in \eqref{Gua EOM} with $m^a_1$, and the second one is the contraction of \eqref{curly P EOM} with $m^a_1$. In both cases this follows from the fact that $\partial_um^a_1=0$. Let us then rewrite \eqref{Psi0} as
\be
\Psi_0=\f{\Psi_0^{-1}}{r^4}+\f{1}{r^5}\Big(\Psi_0^0+\Psi_0^{0,1}(\ln r)+\Psi_0^{0,2}(\ln r)^2\Big)+\O(r^{-6}).
\ee
Expanding \eqref{NP0} then leads to
\be\label{expNP0}
\!\!\partial_u\Psi_0^{-1}=0=\partial_u\Psi_0^{0,2},
\q
\partial_u\Psi_0^{0,1}=\eth\Psi_1^{0,1},
\q
\partial_u\Psi_0^0=\eth\Psi_1^0-3\sigma_2\Psi_2^0+\left(4\partial_u\epsilon_2-\f{3}{4}R\right)\Psi_0^{-1}.
\ee
Using \eqref{Weyl scalars} and \eqref{covariant functionals}, one can show that these equations are equivalent to the contraction of \eqref{EOM E10}, \eqref{EOM E11} and \eqref{EOM E12} with $m^a_1m^b_1$.

\subsection{Asymptotic symmetries}
\label{sec:AKV}

We now study the asymptotic symmetries of the class of LAF spacetimes defined by \eqref{angular metric}. As expected, since here $q_{ab}$ is completely arbitrary, these symmetries are given by the group BMSW \cite{Freidel:2021fxf}, which is gBMS with the additional freedom of performing Weyl rescalings of $\sq$. It is important to note that this result is obtained from a gauge-fixing approach in the physical spacetime, and that it is therefore insensitive to the loss of peeling and the unavailability of a smooth conformal compactification.

Preserving the Bondi gauge conditions \eqref{Bondi gauge} implies that the asymptotic Killing vector field $\xi=\xi^u\partial_u+\xi^r\partial_r+\xi^a\partial_a$ has temporal and angular components given by
\be\label{AKBondigauge}
\!\!\xi^u=f,
\q
\xi^a=Y^a+I^a=Y^a-\int_r^\infty\de r'\,e^{2B}\gamma^{ab}\partial_bf=Y^a-\f{\partial^af}{r}+\f{C^{ab}\partial_bf}{2r^2}+\O(r^{-3}),
\ee
with $\partial_rf=0=\partial_rY^a$.
In order to preserve the polyhomogeneous expansion \eqref{angular metric} of the angular metric, the radial part of the vector field has to be of the form
\be
\xi^r=-rW+\zeta_0+\f{\zeta_1}{r}+\f{1}{r^2}\big(\zeta_2+\zeta_{2,1}(\ln r)\big)+\f{1}{r^3}\big(\zeta_3+\zeta_{3,1}(\ln r)+\zeta_{3,2}(\ln r)^2\big)+\O(r^{-4}),
\ee
where $W$ parametrizes the Weyl transformations. The expansion of $\xi^r$ can be determined in terms of $U^a$ and $\xi^a$ using the differential determinant condition $\partial_r(\pounds_\xi\ln\gamma)=0$, which implies with $\gamma=r^4q$ and $\partial_uq=0$ that
\be\label{preserving det condition}
\xi^r=-rW+\f{r}{2}(U^a\partial_af-D_a\xi^a).
\ee
In particular, we find
\be
\zeta_0=\f{1}{2}\Delta f,
\q\q
\zeta_1=-\f{1}{2}D_bC^{ab}\partial_bf-\f{1}{4}C^{ab}D_a\partial_bf.
\ee
Finally, preserving the boundary conditions also imposes that $\partial_uf=W$ and $\partial_uY^A=0=\partial_uW$, so that one can write
\be\label{f as T and W}
f(u,x)=T(x)+uW(x),
\ee
where $T$ is a pure supertranslation.

We then have all the necessary ingredients to compute the transformation laws of the various fields appearing in the on-shell line element. Focusing on the transformations which are necessary in order to compute the symplectic flux, we have
\bsub\label{transformation laws}
\be
\delta_\xi R&=\pounds_YR+2(\Delta+R)W,\\
\delta_\xi\ln\sq&=D_aY^a-2W,\label{transformation volume}\\
\delta_\xi q_{ab}&=\big(\pounds_Y-2W\big)q_{ab},\\
\delta_\xi q^{ab}&=\big(\pounds_Y+2W\big)q^{ab},\\
\delta_\xi C_{ab}&=\big(f\partial_u+\pounds_Y-W\big)C_{ab}-2D_{\la a}\partial_{b\ra}f,\\
\delta_\xi C^{ab}&=\big(f\partial_u+\pounds_Y+3W\big)C^{ab}-2D^{\la a}\partial^{b\ra}f,\\
\delta_\xi D_{ab}&=(f\partial_u+\pounds_Y)D_{ab},\\
\delta_\xi N_{ab}&=\big(f\partial_u+\pounds_Y\big)N_{ab}-2D_{\la a}\partial_{b\ra}W,\\
\delta_\xi\M&=\big(f\partial_u+\pounds_Y+3W\big)\M+\J^a\partial_af.
\ee
\esub
One can note that the presence of $D_{ab}$ and the loss of peeling does not affect any of these transformation laws, and that $D_{ab}$ itself transforms homogeneously.

\section{Symplectic structure, charges and fluxes}
\label{sec:charges and all}

In the previous section we have defined and characterized a family of LAF spacetimes. We can now study the associated radiative symplectic structure, the asymptotic charges, and the corresponding fluxes. In particular, we are going to show that although (in the case of gBMS or BMSW) the superrotation charges contain a divergent $(\ln r)$ term sourced by $D_{ab}$, the integrated flux on $\I^+$ is independent of $D_{ab}$.

\subsection{Symplectic structure}
\label{subsec:symplectic}

We start with the symplectic potential descending from the Einstein--Hilbert Lagrangian, whose expression is
\be
\theta^\mu=\sqrt{-g}\big(g^{\alpha\beta}\delta\Gamma^\mu_{\alpha\beta}-g^{\mu\alpha}\delta\Gamma^\beta_{\alpha\beta}\big).
\ee
Computing the $u$ and $r$ components, we find
\be
\theta^u=r\theta^u_\text{div}+\theta^u_0+\f{\theta^u_1}{r}+\O(r^{-2}),
\q\q
\theta^r=r\theta^r_\text{div}+\theta^r_0+\O(r^{-1}),
\ee
where
\bsub\label{theta components u and r}
\be
\theta^u_\text{div}&=2\delta\sq,\\
\theta^u_0&=\f{1}{2}\sq\,\delta q^{ab}C_{ab},\\
\theta^u_1&=\sq\,\delta q^{ab}D_{ab},\\
\theta^r_\text{div}&=-\f{1}{2}\sq\,\delta q^{ab}N_{ab}-\delta\big(\sq\,R\big),\\
\theta^r_0
&=\f{1}{2}\sq\,N_{ab}\delta C^{ab}-\f{1}{4}\sq\,\delta q^{ab}\big(RC_{ab}+2\partial_uD_{ab}\big)-\f{1}{2}\sq\,q^{ab}\delta\big(D_aD^cC_{bc}\big)\cr
&\pe+2\delta\sq\left(\M+\f{3}{32}\partial_u[CC]+\f{1}{4}D_aD_bC^{ab}\right)\cr
&\pe+2\delta\left(\sq\,\M-\f{3}{32}\sq\,\partial_u[CC]\right)-\f{1}{2}\sq\,D_a\Big(\delta\ln\sq\,D_bC^{ab}\Big).
\ee
\esub
It should be noted that in these expressions we have $\delta\sq\neq0$ because for the sake of generality we consider the context of BMSW. This is reflected in the fact that the asymptotic symmetries have a free function $W$ related to the transformation of the volume $\sq$ as in \eqref{transformation volume}.

To obtain from first principles a symplectic structure on $\I^+$ \cite{Ashtekar:1981bq,Ashtekar:1990gc}, we compute the symplectic structure on a constant Cauchy time slice $\Sigma_t$ with $t=u+r$, and then take the limit $t\to+\infty$ while keeping $u$ constant, or equivalently the limit $r\to+\infty$. For the symplectic potential we therefore need to consider the component
\be
\theta^t
&=\theta^u+\theta^r\cr
&=r\theta^t_\text{div}+\theta^t_0+\O(r^{-1})\cr
&=t\theta^t_\text{div}+\big(\theta^t_0-u\theta^t_\text{div}\big)+\O(r^{-1}).
\ee
The $t$-divergent piece in this component is given by
\be
\theta^t_\text{div}=\theta^u_\text{div}+\theta^r_\text{div}=\f{1}{2}\sq\,q^{ab}\delta N_{ab}+\delta\big(\sq\,(2-R)\big),
\ee
while the finite piece is
\be\label{theta t finite}
\theta^t_\text{finite}
&=\theta^t_0-u\theta^t_\text{div}\cr
&=\theta^u_0+\theta^r_0-u\big(\theta^u_\text{div}+\theta^r_\text{div}\big)\cr
&=\f{1}{2}\sq\,N_{ab}\delta C^{ab}-\f{1}{4}\sq\,\delta q^{ab}RC_{ab}-\f{1}{2}\sq\,q^{ab}\delta\big(D_aD^cC_{bc}\big)\cr
&\pe+2\delta\sq\left(\M+\f{3}{32}\partial_u[CC]+\f{1}{4}D_aD_bC^{ab}\right)+\f{1}{2}\partial_u\Big(\sq\,\big(uC_{ab}-D_{ab}\big)\delta q^{ab}\Big)\cr
&\pe+\delta\left(2\sq\,\M-\f{3}{16}\sq\,\partial_u[CC]+u\sq\,(R-2)\right) -\f{1}{2}\sq\,D_a\Big(\delta\ln\sq\,D_bC^{ab}\Big).
\ee
Denoting
\be
\int_{\I^+}\coloneqq\oint_{S^2}\de^2x\int_{-\infty}^{+\infty}\de u,
\ee
we finally find that the symplectic structure has a $t$-divergent piece given by
\be\label{Omega div}
\Omega_{\text{div}}=\int_{\I^+}\delta\theta^t_\text{div}=\f{1}{2}\int_{\I^+}\delta\big(\sq\,q^{ab}\big)\delta N_{ab},
\ee
and a finite piece given by
\be\label{Omega finite}
\Omega
&=\int_{\I^+}\delta\theta^t_\text{finite}\cr
&=\int_{\I^+}\f{1}{2}\delta\big(\sq\,N_{ab}\big)\delta C^{ab}-\f{1}{4}\delta\big(\sq\,RC_{ab}\big)\delta q^{ab}-\f{1}{2}\delta\big(\sq\,q^{ab}\big)\delta\big(D_aD^cC_{bc}\big)\cr
&\pe\phantom{\int_{\I^+}}-2\delta\sq\,\delta\left(\M+\f{3}{32}\partial_u[CC]+\f{1}{4}D_aD_bC^{ab}\right)+\f{1}{2}\partial_u\Big[\delta\Big(\sq\,\big(uC_{ab}-D_{ab}\big)\Big)\delta q^{ab}\Big].\q
\ee
Several comments are in order.
\begin{enumerate}
\item The finite radiative symplectic structure \eqref{Omega finite} is insensitive to the fate of the peeling at $\I^+$. Indeed, one can see on the second line that the only contribution from $D_{ab}$ to $\Omega$ drops by virtue of the equation of motion \eqref{EOM D and Enn1}.
\item Because we have constructed $\Omega$ as the limit of a symplectic structure on a Cauchy slice \cite{Ashtekar:1981bq,Ashtekar:1990gc}, it agrees with the integral of $\delta\theta^r_0$ over $\I^+$ only in the case where $\delta q_{ab}=0$. When $\delta q_{ab}\neq0$, we get instead that
\be\label{Omega rewritten}
\Omega=\int_{\I^+}\delta\theta^r_0-\f{1}{2}\int_{\I^+}\partial_u\Big[\delta q^{ab}\delta\Big(\sq\,uC_{ab}\Big)\Big].
\ee
\item The finite contribution $\Omega$ can be understood as a renormalized symplectic structure on $\I^+$. Indeed, since $\partial_uq_{ab}=0$ we can write the divergent part $\Omega_\text{div}$ as a corner term
\be\label{Omega div corner}
\Omega_{\text{div}}=\f{1}{2}\int_{\I^+}\partial_u\Big[\delta\big(\sq\,q^{ab}\big)\delta C_{ab}\Big].
\ee
The corner ambiguities in the symplectic structure \cite{Compere:2008us,Geiller:2017xad} can then be exploited to subtract this divergent contribution and isolate the finite part $\Omega$.
\item This renormalization was not required in previous work \cite{Campiglia:2015yka}, since there the symplectic structure was restricted to a subspace of the radiative phase space where the shear has large $|u|$ fall-offs given by
\be\label{shear subspace}
C_{ab}(u,x)\big|_{|u|\to\infty}=\O\left(\f{1}{|u|^{1+\epsilon}}\right),\q\epsilon>0.
\ee
Clearly, with these fall-off conditions the contribution $\Omega_{\textrm{div}}$ written as \eqref{Omega div corner} vanishes. For a discussion of weaker fall-offs of the shear compatible with superrotations see \cite{Compere:2018ylh}.
\item Although the renormalized symplectic structure $\Omega$ has no linear divergence thanks to the possibility of removing the corner term $\Omega_{\textrm{div}}$, in the presence of massive sources (i.e. when the mass aspect at $\I^+_+$ does not vanish) $\Omega$ exhibits a logarithmic divergence which cannot be cancelled by corner terms at $\I^+$. In a nutshell, this is due to the fact that when the mass aspect at  $i^+$ is non-zero, the ``true'' retarded coordinate $u$ differs from the flat coordinate $u=t-r$. For example, in the particular case of an isotropic mass aspect $M(u=+\infty,x)=M_+$, the retarded coordinate is modified by a logarithmic term as
\be
u=t-r-2GM_+\ln(r-2M_+).
\ee
Therefore, as $r\to\infty$ with $(u,x)$ fixed, we get $r\to t-u-2GM_+(\ln t)$ with $t\to\infty$. As a result $\theta^t_\text{finite}=\theta_0^t-u\theta_\text{div}^t$ constructed in \eqref{theta t finite} is shifted by an additional term as
\be
\theta^t_\text{finite}\to\theta^t_\text{finite}-2GM_+\theta^r_\text{div}(\ln t).
\ee
The detailed study of this mechanism depends on the matter sources and is performed in \cite{Choi-Laddha-Puhm-WIP}, but we can already deduce the structural properties of such a divergent contribution from our analysis. In particular, the gBMS flux at $\I^+$ is also modified by the presence of this logarithmically-divergent contribution. For a generic scattering involving a massive scalar field, this results in the presence of an additional ``drag contribution'' computed in \cite{Choi-Laddha-Puhm-WIP}. Since the divergent contribution depends on the matter fields at $i^+$, it cannot be cancelled by a purely gravitational counter-term on the celestial sphere at $\I^+_+$, and most likely requires a detailed analysis of the covariant phase space for the coupled gravity-matter system. In the present work we consider for simplicity that the mass aspect at $\I^+_+$ vanishes, so that this subtlety does not arise.
\end{enumerate}

\subsection{Asymptotic charges}

We now compute the bare asymptotic charges on a constant $u$ cut, i.e. the charges which follow from a standard covariant phase space analysis without performing any symplectic renormalization or adding corner terms. In terms of the Komar aspect\footnote{We have $P^{[\mu}Q^{\nu]}=(P^\mu Q^\nu-P^\nu Q^\mu)/2$.} $K_\xi^{\mu\nu}\coloneqq-\sqrt{-g}\,\nabla^{[\mu}\xi^{\nu]}$, these charges are given by the $(ur)$ component of the Iyer--Wald formula \cite{Iyer:1994ys}
\be
\slashed{\delta}Q_\xi^{\mu\nu}
&=2\big(\delta K^{\mu\nu}_\xi-K^{\mu\nu}_{\delta\xi}+\xi^{[\mu}\theta^{\nu]}\big)\cr
&=2\sqrt{-g}\left[\xi^{[\mu}\big(\nabla^{\nu]}\delta g-\nabla_\alpha\delta g^{\nu]\alpha}\big)+\xi_\alpha\nabla^{[\mu}\delta g^{\nu]\alpha}+\left(\f{1}{2}\delta gg^{[\mu\alpha}-\delta g^{[\mu\alpha}\right)\nabla_\alpha\xi^{\nu]}\right],
\ee
where the variations are $\delta g^{\mu\nu}=\delta(g^{\mu\nu})$ and $\delta g=g_{\mu\nu}\delta g^{\mu\nu}=-\delta\ln g$. For the component of interest we find the expansion
\be
\slashed{\delta}Q_\xi^{ur}=r\slashed{\delta}Q_\text{div}+(\ln r)\slashed{\delta}Q_\text{ln-div}+\slashed{\delta}Q_Y+\slashed{\delta}Q_W+\slashed{\delta}Q_f+\O(r^{-1}),
\ee
with
\bsub\label{bare charge}
\be
\slashed{\delta}Q_\text{div}&=\f{1}{2}f\Big(\sq\,N^{ab}\delta q_{ab}-2\delta\big(\sq\,R\big)\Big)-\Delta f\delta\sq-\f{1}{2}W\sq\,q^{ab}\delta C_{ab}+\delta\Big(\sq\,Y^aD^bC_{ab}\Big),\label{Q div}\\
\slashed{\delta}Q_\text{ln-div}&=2\delta\Big(\sq\,Y^aD^bD_{ab}\Big),\label{log r IW charge}\\
\slashed{\delta}Q_Y&=2\delta\bigg(\sq\,Y^a\left[\P_a-\f{1}{4}C_{ab}D_cC^{bc}-\f{3}{32}\partial_a[CC]+D^bD_{ab}\right]\bigg),\\
\slashed{\delta}Q_W&=-\f{1}{8}\delta\Big(\sq\,W[CC]\Big)-W\sq\,q^{ab}\delta D_{ab},\\
\slashed{\delta}Q_f
&=4f\delta\big(\sq\,\M\big)-\f{1}{2}f\sq\,C_{ab}\delta N^{ab}\cr
&\pe+\sq\,\delta q_{ab}\left[f\left(\f{1}{4}RC^{ab}-\f{1}{2}D^aD_cC^{cb}+\f{1}{2}\partial_uD^{ab}\right)-\partial^afD_cC^{cb}+\f{1}{4}C^{ab}\Delta f\right]\cr
&\pe+\delta\sq\,\left[f\left(2\M-\f{9}{16}\partial_u[CC]+D_aD_bC^{ab}\right)+\f{3}{4}C^{ab}D_a\partial_bf+2D_bC^{ab}\partial_af\right]\cr
&\pe-\f{1}{2}\sq\,D_a\Big(f\delta\big(D_bC^{ab}\big)+2fD_bC^{ab}\delta\ln\sq\Big).
\ee
\esub
Several observations should be made about these charges.
\begin{enumerate}
\item We have written here the charge aspects, and not the integrated charges over the 2-sphere at a constant $u$ cut.
\item The tensor $D_{ab}$ responsible for the loss of peeling appears explicitly in the charges, and in particular gives rise to a logarithmically-divergent term. Moreover, this divergent term is only non-vanishing in the case of gBMS (or BMSW). Indeed, when fixing $\delta q_{ab}=0$ we have to impose $2D_{(a}Y_{b)}=q_{ab}(D_cY^c)$, which then implies the vanishing of \eqref{log r IW charge} after integration by parts since $D_{ab}$ is trace-free.
\item The charges also contain a linearly-divergent contribution due to the fact that $\delta q_{ab}\neq0$.
\item These two divergent contributions can be renormalized using the technique of symplectic renormalization \cite{Compere:2018ylh,Freidel:2019ohg,McNees:2023tus,Freidel:2024tpl,McNees:2024iyu}. This relies on the use of corner terms which can systematically be identified from the symplectic potential. We demonstrate this explicitly in appendix \ref{app:renormalization}. The interesting novelty of the present situation is due to the logarithmically-divergent term, which can be renormalized using a corner term inferred from the $1/r$ component of $\theta^u$.
\item The charges have been computed for the sake of generality in the framework of BMSW, which explains the presence of a component associated with the symmetry parameter $W$. When reducing BMSW to gBMS by fixing $\delta\sq=0$, this parameter becomes $2W=D_aY^a$ by virtue of \eqref{transformation volume}, and therefore gives a new contribution to the superrotation charges.
\end{enumerate}

\subsection[Flux at $\I^+$]{Flux at $\boldsymbol{\I^+}$}

We now study the flux associated with $\text{Diff}(S^{2})$ superrotations at $\I^+$. In principle, we could compute the full BMSW flux by contracting the symplectic structure \eqref{Omega finite} with the transformation generated by the asymptotic Killing vector $\xi$. This would require in particular to decompose the shear as $C_{ab}=\hat{C}_{ab}+u\rho_{ab}$, where $\rho_{ab}$ is the trace-free part of the Geroch tensor uniquely determined by $\delta_Yq_{ab}$ and such that $\hat{N}_{ab}=\partial_u\hat{C}_{ab}$ transforms homogeneously \cite{Geroch1977,Compere:2018ylh,Campiglia:2020qvc,Rignon-Bret:2024gcx}. This is what was done in \cite{Donnay:2022hkf,Agrawal:2023zea}. However, it will be sufficient for our purposes\footnote{As shown in \cite{Campiglia:2020qvc,Rignon-Bret:2024gcx}, demanding that the charge algebra closes without a 2-cocycle requires a careful treatment of corner terms involving the Geroch tensor $\rho_{ab}$, and affecting the charges and the fluxes. We believe that the analysis in these references will however not be impacted by the presence of $D_{ab}$.} to initially allow for BMSW transformations (so as to probe the effect of non-trivial superrotations), but then project the flux onto a Bondi frame, defined by a choice $q_{ab}=q^\circ_{ab}$ of round sphere metric and by $\rho_{ab}=0$. The flux projected to a Bondi frame is then\footnote{Here $\xi\ipp$ is the contraction in field space between a variational vector field $\delta_\xi$ and a variational 1-form. For example $\xi\ipp(\delta\theta[\delta])=\delta_\xi\theta[\delta]-\delta\theta[\delta_\xi]$. The notations $\xi\ipp$ and $\xi\ip$ are often denoted by $I_\xi$ and $\iota_\xi$.}
\be
\slashed{\delta}\F_\xi\coloneqq(\xi\ipp\Omega)\big|_{q_{ab}\,=\,q_{ab}^\circ},
\ee
where $\xi$ is a generic linear combination of the asymptotic symmetry transformations. Using the transformation laws \eqref{transformation laws}, we find
\be\label{flux Bondi frame}
\slashed{\delta}\F_\xi
&=\int_{\I^+}\sq\,\partial_u\left[\f{1}{2}fN_{ab}\delta C^{ab}+\delta\Big(C_{ab}D^a\partial^bf+\big(uC_{ab}-D_{ab}\big)D^aY^b\Big)\right]\cr
&\pe\phantom{\int_{\I^+}}-\f{1}{2}\delta\Big[\sq\,N_{ab}\Big(fN^{ab}+\pounds_YC^{ab}+3WC^{ab}\Big)\Big]\cr
&\pe\phantom{\int_{\I^+}}+\f{1}{2}\delta\Big[\sq\,C_{ab}\Big(D^a\Delta Y^b+D^aD_cD^bY^c-RD^aY^b-3D^aD^bD_cY^c\Big)\Big]\cr
&\pe\phantom{\int_{\I^+}}-2\delta\left[\sq\,\big(D_aY^a-2W\big)\left(\M+\f{3}{32}\partial_u[CC]-\f{1}{4}D_aD_bC^{ab}\right)\right].
\ee
As a consistency check for this formula, we can verify with a lengthy calculation
that this flux can be rewritten as\footnote{For the divergent part of the symplectic structure one can also show that
\be
\xi\ipp\Omega_\text{div}=\f{1}{2}\int_{\I^+}\partial_u\Big[\delta_\xi\big(\sq\,q^{ab}\big)\delta C_{ab}-\delta\big(\sq\,q^{ab}\big)\delta_\xi C_{ab}\Big]=\int_{\I^+}\partial_u\big(\slashed{\delta}Q_\text{div}\big).
\ee
This follows immediately from the rewriting \eqref{Omega div corner} of the divergent symplectic structure as a corner term, and from the result \eqref{Qdiv from corner term} relating this corner term to the $r$-divergent part of the charges.}
\be
\slashed{\delta}\F_\xi=\int_{\I^+}\partial_u\Big[\slashed{\delta}Q_f+\slashed{\delta}Q_W+\slashed{\delta}Q_Y\Big]\Big|_{q_{ab}\,=\,q_{ab}^\circ}+\int_{\I^+}\partial_u\delta\Big(\sq\,uC_{ab}D^aY^b\Big).
\ee
As expected, this relates the symplectic flux to the time evolution of the charges \eqref{bare charge}, and also reflects the decomposition \eqref{Omega rewritten} of the symplectic structure, as the two terms in this rewriting of the flux arise respectively from the two contributions in \eqref{Omega rewritten}.

The flux formula \eqref{flux Bondi frame} is one of the central results of the paper, so we now discuss its key properties. First, one can note that the flux at $\I^+$ is independent of $D_{ab}$ by virtue of the first equation of motion in \eqref{EOM D and Enn1}. This means in particular that the flux is a priori insensitive to the fact that we are considering a LAF solution space which does not satisfy peeling. However, as explained in the introduction (and exemplified by Christodoulou's argument in appendix \ref{Christo argument}), one should recall that here there is an important physical input which is motivating and sourcing the loss of peeling, namely the $1/|u|$ tail to the memory term present in the shear. Although the flux does not depend on $D_{ab}$, it acquires an important logarithmically-divergent term due to this tail to the memory. We discuss this term below in section \ref{sec:log flux}. We recall that although the present analysis leading to the result \eqref{flux Bondi frame} now includes the contribution from $D_{ab}$ due to the absence of peeling, it disregards the contribution from the Geroch tensor which was carefully discussed in  \cite{Donnay:2022hkf,Agrawal:2023zea}.

Since we are in particular interested in superrotations in order to illustrate these features, let us now reduce\footnote{Weyl transformations provide no new independent conservation laws in addition to the supertranslations, and the flux of $Q_W$ contains no soft contribution.} BMSW to gBMS by setting $2W=D_aY^a$, and then consider trivial supertranslations $T=0$. Recalling that the parameter $f$ decomposes as \eqref{f as T and W} and using the equation of motion $\partial_uD_{ab}=0$, we find that the soft part of the superrotation flux is integrable and given by $\delta\F_{Y,\text{soft}}$ with
\be\label{soft flux}
\F_{Y,\text{soft}}=\f{1}{2}\int_{\I^+}\sq\,\Big[\partial_u\big(uC_{ab}M^{ab}[Y]\big)+C_{ab}O^{ab}[Y]\Big],
\ee
where we have introduced the two symmetric and trace-free tensors
\bsub
\be
M^{ab}[Y]&\coloneqq D^{\la a}D^{b\ra}D_cY^c+2D^{\la a}Y^{b\ra},\\
O^{ab}[Y]&\coloneqq D^{\la a}\Delta Y^{b\ra}+D^{\la a}D_cD^{b\ra}Y^c-RD^{\la a}Y^{b\ra}-3D^{\la a}D^{b\ra}D_cY^c.
\ee
\esub
Let us then consider the asymptotic parametrization of the shear towards the corners of future null infinity, which is
\be\label{general C falloffs}
C_{ab}(u,x)\big|_{u\to\pm\infty}=C^{0,\pm}_{ab}(x)+\f{1}{|u|}C^{1,\pm}_{ab}(x)+\mathscr{C}_{ab}(u,x),
\q\q
\mathscr{C}_{ab}(u,x)=\O\left(\f{\ln u}{u^2}\right).
\ee
Here $C^{0,\pm}_{ab}$ gives rise to the displacement memory, $C^{1,\pm}_{ab}$ is the tail to the memory which is included for consistency with the logarithmic soft graviton theorem, and $\mathscr{C}_{ab}$ contains the rest of the expansion. The combination $C^{0,+}_{ab}+C^{0,-}_{ab}$ is the constant shear mode, while $C^{0,+}_{ab}-C^{0,-}_{ab}$ is the soft news. Below we will also denote the constant shear mode at $\I^\pm$ by $C^0_{ab}\big|_{\I^\pm}$.

Due to the presence of $C^{0,\pm}_{ab}$ and $C^{1,\pm}_{ab}$ in the asymptotic parametrization of the shear, it is clear that the soft flux will contain linearly and logarithmically-divergent contributions in addition to its finite part. Plugging \eqref{general C falloffs} into \eqref{soft flux}, we can work out these various contributions and define a regularized flux as
\be\label{flux rewritten}
\F^\text{reg}_{Y,\text{soft}}(\lambda)\coloneqq\F^\text{finite}_{Y,\text{soft}}(\lambda_0)+\F^\text{lin-div}_{Y,\text{soft}}(\lambda)+\F^\text{log-div}_{Y,\text{soft}}(\lambda),
\ee
where
\bsub
\be
\!\!\F^\text{finite}_{Y,\text{soft}}(\lambda_0)&=\f{1}{2}\oint_{S^2}\sq\,\big(C^{1,+}_{ab}+C^{1,-}_{ab}\big)M^{ab}[Y]+\f{1}{2}\int_{\I^+}\sq\,\mathscr{C}_{ab}O^{ab}[Y]-\F^\text{log}_{Y,\text{soft}}(\lambda_0),\label{finite flux}\\
\!\!\F^\text{log}_{Y,\text{soft}}(\lambda_0)&=\f{1}{2}(\ln\lambda_0)\oint_{S^2}\sq\,\big(C^{1,+}_{ab}-C^{1,-}_{ab}\big)O^{ab}[Y],\label{log flux}\\
\!\!\F^\text{log-div}_{Y,\text{soft}}(\lambda)&=\,\f{1}{2}(\ln\lambda)\;\oint_{S^2}\sq\,\big(C^{1,+}_{ab}-C^{1,-}_{ab}\big)O^{ab}[Y],\label{log div flux}\\
\!\!\F^\text{lin-div}_{Y,\text{soft}}(\lambda)&=\f{\lambda}{2}\oint_{S^2}\sq\,\big(C^{0,+}_{ab}+C^{0,-}_{ab}\big)\big(M^{ab}+O^{ab}\big)[Y]\eqqcolon\f{\lambda}{2}\oint_{S^2}\sq\,C^0_{ab}\big|_{\I^+}\big(M^{ab}+O^{ab}\big)[Y].\label{lin div flux}
\ee
\esub
The last two terms are respectively the logarithmically and linearly-divergent contributions in the limit $\lambda\to\infty$, and the finite scale $\lambda_0$ appears because the logarithmic term must be of the form $\ln(\lambda/\lambda_0)$ for dimensional reasons.

\subsubsection{Linearly-divergent flux}

In \cite{Campiglia:2015yka}, the linearly-divergent term \eqref{lin div flux} is absent because the flux is evaluated on a subspace of the radiative phase space where the shear satisfies \eqref{shear subspace}. However, these fall-off conditions are too restrictive to describe certain important physical situations, and a static shear mode $C_{ab}^{0,\pm}$ is present in the generic shear data at ${\cal I}^{+}$ obtained by gravitational scattering of massive sources\footnote{For example, if we consider $n$ incoming particles at $i^-$ with momenta $p_1,\dots,p_n$, then as shown in \cite{Damour:2020tta} the static mode at $\I^+$ is simply $C^0_{ab}(x)=\sum_{i=1}^n\f{\epsilon_{ab}^{\mu\nu}p_i^\mu p_i^\nu}{p_i \cdot\hat{n}}$, where $\hat{n}\coloneqq(1,x)$ and where the polarization tensor is indexed by two helicity states which can be parametrized in terms of symmetric trace-free tensor on $S^2$.} \cite{Damour:2020tta}. By focusing solely on the flux at $\I^+$, the linearly-divergent term \eqref{lin div flux} must therefore be dealt with. One strategy to do so would be to subtract a counter-term, which must be chosen such that the resulting finite charge generates a closed gBMS charge algebra. Whether such a renormalization is achievable is an open question which remains outside the scope of the paper. This subtlety is also discussed in \cite{Compere:2018ylh}.

On the other hand, as far as the conservation laws are concerned, the divergent term \eqref{lin div flux} on $\I^+$ is not problematic because it can be unambiguously cancelled against the analogous term on $\I^-$. Indeed, at any finite cut-off $\lambda$ the divergent contribution at $\I^\pm$ is given by
\be
\F^\text{lin-div}_{Y,\text{soft}}\big|_{\I^\pm}(\lambda)=\f{\lambda}{2}\oint_{S^2}\sq\,C^0_{ab}\big|_{\I^\pm}\big(M^{ab}+O^{ab}\big)[Y^\pm],
\ee
and assuming the antipodal identification of $Y^\pm$ between $\I^+$ and $\I^-$, as argued in \cite{He:2014laa}, these two flux contributions match. More precisely, this follows from the fact that the constant shear modes are given by\footnote{This is also called the electricity condition, and it follows from the leading classical soft graviton theorem.}
\be
C^0_{ab}\big|_{\I^\pm}=D_{\la a}D_{b\ra}T^\pm,
\ee
where $T^\pm$ are supertranslation generators at $\I^{\pm}$. Since these generators $T^\pm$ are antipodally identified in $D=4$ dimensions \cite{Prabhu:2019fsp,Prabhu:2021cgk} (see also \cite{Campiglia:2017mua} for the analogous result in QED), we get an antipodal identification of the constant shear modes at $\I^+$ and $\I^-$. This therefore implies that \emph{the difference} $\F_Y\big|_{\I^+}-\F_Y\big|_{\I^-}$ between the superrotation fluxes has no linear divergence. Since it is this difference of fluxes which is relevant for the conservation laws associated to classical scattering, we can drop the linearly-divergent contribution $\F^\text{lin-div}_{Y,\text{soft}}$. 

\subsubsection{Logarithmically-divergent flux}
\label{sec:log flux}

If the fall-offs of the shear include the tail to the memory contributions $C^{1,\pm}_{ab}$, as is the case here with our parametrization \eqref{general C falloffs}, the soft charge is logarithmically-divergent. We have here regulated this divergence to obtain the logarithmic contributions \eqref{log flux} and \eqref{log div flux}. Important observations can be made about these contributions.
\begin{enumerate}
\item The integral over $u$ of the terms in $1/|u|$ in \eqref{general C falloffs}, i.e. the tail to the memory, has produced a logarithmic contribution to the soft flux. This contribution is logarithmically-divergent, but the coefficient of $\ln\lambda$ in $\F^\text{log-div}_{Y,\text{soft}}$ is universal. It matches precisely the contribution found in perturbative gravity in \cite{Choi:2024ygx}, with $\lambda$ playing the role of an infrared scale in frequency space.
\item In addition to the divergent contribution, the logarithmic term also brings an ambiguity to the finite part of the soft charge. This is clear from the presence of $\ln\lambda_0$, where $\lambda_0$ is any finite scale determined by the scattering configuration\footnote{In a typical scattering configuration involving two objects, $\lambda_0\sim|\vec{b}|$ where $\vec{b}$ is the impact parameter.} and cannot be specified a priori.
\item In the finite part \eqref{finite flux} of the soft charge, the piece arising from $\mathscr{C}_{ab}$ is the ``usual'' contribution present with the boundary conditions \eqref{shear subspace} (i.e. in the absence of the infrared tail to the memory), which gives rise to the spin memory upon looking at the conservation laws. In addition to the ambiguity brought by $\lambda_0$, one can see here that the finite part also receives a contribution from $C^{1,\pm}_{ab}$, which therefore modifies the spin memory formula. The ``usual'' spin memory formula \cite{Pasterski:2015tva}, which is derived ignoring the tail terms in the shear and assuming peeling, is given for a generic scattering by
\be
\F^\circ_{Y,\text{soft}}\big|_{\I^-}-\F^\circ_{Y,\text{soft}}\big|_{\I^+}=\big(\F_{Y,\text{hard}}\big|_{\I^+}+\F_{Y,\text{matter}}\big|_{i^+}\big)-\big(\F_{Y,\text{hard}}\big|_{\I^-}+\F_{Y,\text{matter}}\big|_{i^-}\big).
\ee
Here, due to the presence of the tail terms, this formula is modified to
\be\label{new spin memory}
\F_{Y,\text{soft}}\big|_{\I^+}-\F_{Y,\text{soft}}\big|_{\I^-}
&=\;\F^\circ_{Y,\text{soft}}\big|_{\I^+}-\F^\circ_{Y,\text{soft}}\big|_{\I^-}\cr
&\pe\;+\f{1}{2}\oint_{S^2}\sq\,\triangle\big(C^{1,+}_{ab}+C^{1,-}_{ab}\big)M^{ab}[Y]+\O(\ln\lambda_{0}),\q
\ee
where
\be
\triangle C^{1,+}_{ab}=C^{1,+}_{ab}\big|_{\I^+}-C^{1,+}_{ab}\big|_{\I^-}.
\ee
This therefore shows that the tail to the memory affects the spin memory formula in $D=4$ dimensions, both by bringing a new finite contribution to the soft charge and by introducing a regularization ambiguity due to the logarithmic term.
\end{enumerate}

In summary, we have computed the soft superrotation flux for LAF spacetimes of the type \eqref{angular metric} and in the presence of the tail to the memory predicted by the classical soft theorem. Our findings confirm the results of 
\cite{Choi:2024ygx} (see equations (27) and (28) there). In this reference however, the gravitational contribution to the flux was assumed to be of the same form as the gBMS flux derived from the solution space satisfying peeling, i.e. with $D_{ab}=0=E^{n,m>0}_{ab}$. Here we have justified why this flux formula holds even with the loss of peeling caused by the tail to the memory. We have also analyzed and regularized the different divergent contributions to the flux.  

Using an antipodal identification between $\I^+_-$ and $\I^-_+$, as well as the ``logarithmically divergent'' matter symplectic structure at $i^\pm$, the conservation law for the total (i.e. gravitational and matter contribution) superrotation flux has been shown to be equivalent to the classical logarithmic soft graviton theorem in \cite{Choi:2024ygx,Choi-Laddha-Puhm-WIP}. The relationship between quantum superrotation charges and the so-called quantum logarithmic soft theorem derived by Sahoo and Sen has been analyzed in \cite{Agrawal:2023zea}.

\section{Outlook}

Although the peeling property plays a central role in the notion of asymptotic flatness \cite{Penrose:1962ij,Penrose:1965am,Geroch1977,Sachs:1961zz}, there have been numerous indications over the years that this criterion might not be satisfied for sufficiently physically-relevant situations \cite{Friedrich:1983vx,PhysRevD.19.3483,1985FoPh...15..605W,PhysRevD.19.3495,1986mgm..conf..365D,doi:10.1098/rspa.1981.0101,2002nmgm.meet...44C,Andersson:1993we,Valiente-Kroon:2002xys,Kroon:2004me,Kehrberger:2021uvf,Kehrberger:2021vhp,Kehrberger:2021azo,Kehrberger:2024clh,Kehrberger:2024aak,Gajic:2022pst,Kehrberger:2023btg,Bieri:2023cyn}. Confirming earlier insights by e.g. Damour and Christodoulou, the classical soft graviton theorem of Saha, Sahoo and Sen \cite{Sahoo:2018lxl,Saha:2019tub} has recently confirmed the ubiquitous presence of universal tail to the memory contributions in the radiative metric for a generic gravitational scattering. The first of these contributions is the $1/|u|$ mode related to the leading logarithmic soft graviton theorem, and the presence of this tail mode confirms the systematic loss of peeling in a generic scattering.

In light of this observation, it is therefore clear that the standard Bondi--Sachs arena for describing asymptotic-flatness and asymptotic symmetries \cite{Bondi:1960jsa,Bondi:1962px,Bondi:1962rkt,Sachs:1961zz,Sachs:1962wk,Newman:1961qr} must be extended. The way to do so has been known for quite some time using the framework of polyhomogeneous expansions \cite{1985FoPh...15..605W,Chrusciel:1993hx,Kroon:1998tu,ValienteKroon:2002gb}, although this had not been used to systematically analyse the asymptotic symmetries and the charges. In the present paper we have closed this gap by studying the detailed asymptotic structure of logarithmically-asymptotically-flat (LAF) spacetimes with line element given by \eqref{Bondi gauge} and \eqref{angular metric}. In section \ref{lafspacetimes} we have studied the solution space and the evolution equations on $\I^+$. This has in particular revealed the structure of a new tower of evolution equations for the modes $E^{n,m>0}_{ab}$ in the expansion \eqref{angular metric}. We have then shown in section \ref{sec:AKV} that this LAF solution space is preserved under the action of the whole BMSW group of asymptotic symmetries, just as in the absence of peeling. In section \ref{sec:charges and all} we have then computed the asymptotic codimension-two charges, and shown that they are sensitive to the loss of peeling when the leading boundary metric is such that $\delta q_{ab}\neq0$. Using the symplectic structure constructed from first principles, we have then compute the flux of the gBMS charges in a Bondi frame, and in particular studied the soft superrotation flux. This flux has been shown to be divergent due to the tail mode, which upon regularization reproduces the classical logarithmic soft graviton theorem as shown in \cite{Choi:2024ygx,Choi-Laddha-Puhm-WIP}. Our flux formula also justifies consistently the flux used in \cite{Agrawal:2023zea}, although in this reference the contribution of the Geroch tensor was also included (while however still assuming peeling).

For future work, it would be interesting to investigate how the solution space which we have constructed could be used to describe flux-balance laws and memory effects arising from e.g. hyperbolic encounters \cite{PhysRevD.1.1559,PhysRevD.19.3483,PhysRevD.19.3495,1978ApJ...224...62K,Hait:2022ukn,Bini:2023gaj}, which we expect to also lead to a loss of peeling. An important (and somehow related) question concerns the role of the tower of subleading logarithmic evolution equations studied in section \ref{sec:log fb laws}, and whether they are related to subleading logarithmic soft graviton theorems. It is tantalizing to speculate that the presence of the first mode in this tower, namely $E_{ab}^{1,2}$, is tied to the subleading classical logarithmic soft theorem which was conjectured by Saha Sahoo and Sen in \cite{Saha:2019tub} and later proved by Sahoo in \cite{Sahoo:2020ryf}. Looking even further ahead, it would be incredibly interesting to see if the analysis of higher modes can shed light on the sub$^{n\geq3}$-leading logarithmic soft theorems which are conjectured to be universal. Finally, this then begs the question of whether this also leads to a new symmetry structure as the $w_{1+\infty}$ symmetry found in the ``standard'' peeling solution space from the evolution equation of $E^{n,0}_{ab}$ \cite{Freidel:2021ytz,Compere:2022zdz,Geiller:2024bgf,Kmec:2024nmu}.

\section*{Acknowledgements}

We would like to thank Sayali Bhatkar, Miguel Campiglia, Sangmin Choi, Piotr Chru\'sciel, Laurent Freidel, Oscar Fuentealba, Marc Henneaux, Leohnard Kehrgerber, Andrea Puhm, Romain Ruzziconi, Biswajit Sahoo, Ashoke Sen and Tom Wetzstein for discussions. AL is especially indebted to Miguel Campiglia, Sangmin Choi, Andrea Puhm and Ashoke Sen for insightful discussions and collaborations on related projects. Research at Perimeter Institute is supported in part by the Government of Canada through the Department of Innovation, Science and Economic Development Canada and by the Province of Ontario through the Ministry of Colleges and Universities.

\appendix

%

\section{Details about the polyhomogeneous solution space}

In this appendix we gather various lengthy expressions appearing in the radial expansion of the polyhomogeneous solution space.

\subsection{Expansion of the angular metric}
\label{app:angular expansion}

The explicit expansion of the angular metric \eqref{angular metric} to order $r^{-3}$ is given by
\be
\gamma_{ab}
&=r^2q_{ab}+rC_{ab}+r\left\{D_{ab}+\f{1}{4}q_{ab}[CC]\right\}\cr
&\hspace{2.48cm}+\f{1}{r}\left\{E^{1,0}_{ab}+\f{1}{2}q_{ab}[CD]+E^{1,1}_{ab}(\ln r)+E^{1,2}_{ab}(\ln r)^2\right\}\cr
&\hspace{2.39cm}+\f{1}{r^2}\bigg\{E^{2,0}_{ab}+\f{1}{2}q_{ab}\left([CE^{1,0}]+\f{1}{2}[DD]-\f{1}{16}[CC]^2\right)\cr
&\hspace{5.1cm}+\left(E^{2,1}_{ab}+\f{1}{2}q_{ab}[CE^{1,1}]\right)(\ln r)\cr
&\hspace{5.1cm}+\left(E^{2,2}_{ab}+\f{1}{2}q_{ab}[CE^{1,2}]\right)(\ln r)^2+E^{2,3}_{ab}(\ln r)^3\bigg\}\cr
&\hspace{2.39cm}+\f{1}{r^3}\bigg\{E^{3,0}_{ab}+\f{1}{2}q_{ab}\left([CE^{2,0}]+[DE^{1,0}]-\f{1}{4}[CC][CD]\right)\cr
&\hspace{5.1cm}+\left(E^{3,1}_{ab}+\f{1}{2}q_{ab}\big([CE^{2,1}]+[DE^{1,1}]\big)\right)(\ln r)\cr
&\hspace{5.1cm}+\left(E^{3,2}_{ab}+\f{1}{2}q_{ab}\big([CE^{2,2}]+[DE^{1,2}]\big)\right)(\ln r)^2\cr
&\hspace{5.1cm}+\left(E^{3,3}_{ab}+\f{1}{2}q_{ab}[CE^{2,3}]\right)(\ln r)^3+E^{3,4}_{ab}(\ln r)^4\bigg\}\q\q\cr
&\hspace{2.39cm}+\O(r^{-4}),
\ee
where it is understood in the notation that $\O(r^{-4})$ contains also logarithmic terms.

\subsection{Expansion of the spacetime metric}
\label{app:metric expansion}

Here we gather the first few terms appearing in the on-shell expansion \eqref{expansions of BUV} of the metric functions $B$, $U^a$ and $V$. More precisely, we go to $n=4$ for $B$, to $n=5$ for $U^a$, and $n=2$ for $V$, but omit to write some lengthier expressions which are not needed for the flux-balance laws given in appendix \ref{app:angular Einstein expansion}. Solving the hypersurface equations $G_{rr}=G_{ra}=G_{ru}=0$ with \eqref{expansions of BUV}, we find
\be
B_{4,0}&=\f{1}{128}\left([CC]^2-12[CE^{1,0}]-8[DD]+[CE^{1,1}]+\f{1}{2}[CE^{1,2}]\right),\\
B_{4,1}&=-\f{3}{32}[CE^{1,1}]+\f{1}{64}[CE^{1,2}],\\
B_{4,2}&=-\f{3}{32}[CE^{1,2}],\\
U^a_{3,1}&=-\f{2}{3}D_bD^{ab},\\
U^a_{4,0}
&=-\f{3}{4}C^{ab}N_b-\f{1}{16}C^{ab}\partial_b[CC]+\f{1}{64}[CC]D_bC^{ab}+\f{5}{24}C^{ab}D^cD_{bc}+\f{1}{8}D^{bc}D^aC_{bc}\cr
&\pe-\f{1}{24}\partial^a[CD]+\f{3}{4}D_bE^{ab}_{1,0}+\f{11}{16}D_bE^{ab}_{1,1}+\f{43}{32}D_bE^{ab}_{1,2},\\
U^a_{4,1}&=\f{1}{2}C^{ab}D^cD_{bc}+\f{3}{4}D_bE^{ab}_{1,1}+\f{11}{8}D_bE^{ab}_{1,2},\\
U^a_{4,2}&=\f{3}{4}D_bE^{ab}_{1,2},\\
U^a_{5,0}&=(\text{lengthy}),\\
U^a_{5,1}&=(\text{lengthy}),\\
U^a_{5,2}&=\f{27}{50}D_bE^{ab}_{2,3}+\f{2}{5}D_bE^{ab}_{2,2}-\f{2}{5}C^{ab}D^cE^{1,2}_{bc}+\f{1}{16}\partial^a[CE^{1,2}]-\f{1}{10}C^{bc}D^aE^{1,2}_{bc},\\
U^a_{5,3}&=\f{2}{5}D_bE^{ab}_{2,3},\\
V_{1,0}&=\f{1}{2}\left(D_aN^a-\f{1}{3}D_aD_bD^{ab}-(D_aC^{ab})(D^cC_{cb})-\f{1}{2}C^{ab}D_aD^cC_{bc}-\f{1}{16}(\Delta+R)[CC]\right),\\
V_{1,1}&=-\f{1}{3}D_aD_bD^{ab},\\
V_{2,0}&=(\text{lengthy}),\\
V_{2,1}&=\f{5}{8}D_aD_bE^{ab}_{1,2}+\f{1}{4}D_aD_bE^{ab}_{1,1}-\f{1}{2}(D_aC^{ab})(D^cD_{cb}),\\
V_{2,2}&=\f{1}{4}D_aD_bE^{ab}_{1,2}.
\ee
With this expansion the order at which the Einstein equations are solved is
\be
G_{rr}=\O(r^{-7}),
\q\q
G_{ra}=\O(r^{-6}),
\q\q
G_{ru}=\O(r^{-6}),
\ee
where we recall that $\O(r^{-n})$ also implicitly contains terms of the form $r^{-n}(\ln r)^m$.

\subsection{Expansion of the angular equations}
\label{app:angular Einstein expansion}

Here we give the explicit form of some of the evolution equations appearing in the expansion \eqref{angular EOMs compact}. Up to $n=4$ we find
\bsub\label{angular EOMs}
\be
\partial_uD_{ab}&=0,\label{Dab EOM}\\
\partial_uE^{1,0}_{ab}&=\f{5}{36}\left(\Delta-\f{14}{5}R\right)D_{ab}+\f{1}{3}D_{\la a}\P_{b\ra}+\f{1}{2}\big(C_{ab}\M+\widetilde{C}_{ab}\widetilde{\M}\big)+\f{1}{16}\partial_u\big([CC]C_{ab}\big),\label{EOM E10}\\
\partial_uE^{1,1}_{ab}&=\f{1}{6}\big(\Delta-R\big)D_{ab},\label{EOM E11}\\
\partial_uE^{1,2}_{ab}&=0,\label{EOM E12}\\
\partial_uE^{2,0}_{ab}&=(\text{lengthy}),\label{EOM E20}\\
\partial_uE^{2,1}_{ab}
&=-\f{1}{4}\big(\Delta+R\big)E^{1,1}_{ab}-\f{3}{8}\left(\Delta-\f{7}{3}R\right)E^{1,2}_{ab}\cr
&\pe-\f{1}{6}C_{ab}D_cD_dD^{cd}+\f{1}{6}C^{cd}\big(D_{\langle a}D_{b\rangle}D_{cd}-D_cD_dD_{ab}\big)
-\f{1}{3}D^cC_{\langle ac}D^dD_{b\rangle d},\label{EOM 21}\\
\partial_uE^{2,2}_{ab}&=-\f{1}{4}\big(\Delta+R\big)E^{1,2}_{ab},\\
\partial_uE^{2,3}_{ab}&=0,\\
\partial_uE^{3,0}_{ab}&=(\text{lengthy}),\label{EOM E30}\\
\partial_uE^{3,1}_{ab}&=(\text{lengthy}),\\
\partial_uE^{3,2}_{ab}
&=-\f{33}{200}\left(\Delta-\f{61}{11}R\right)E^{2,3}_{ab}-\f{3}{20}\big(\Delta+4R\big)E^{2,2}_{ab}+\f{9}{4}\Big(E^{1,2}_{ab}\M+\widetilde{E}^{1,2}_{ab}\widetilde{\M}\Big)\cr
&\pe-\f{1}{2}D^cD_{\langle ac}D^dD_{b\rangle d}+\f{3}{10}D^cC_{\langle ac}D^dE^{1,2}_{b\rangle d}+\f{9}{20}D_cC^{cd}D_dE^{1,2}_{ab}\cr
&\pe+\f{3}{8}C^{cd}D_cD_dE^{1,2}_{ab}-\f{3}{10}C^{cd}D_cD_{\la a}E^{1,2}_{b\ra d}+\f{3}{8}E^{cd}_{1,2}D_cD_dC_{ab}-\f{3}{8}E^{cd}_{1,2}D_cD_{\la a}C_{b\ra d}\cr
&\pe+\f{3}{8}N_{ab}[CE^{1,2}]-\f{1}{8}E^{1,2}_{ab}\partial_u[CC]+\f{3}{16}E^{1,2}_{\la ac}N^{cd}C_{db\ra},\label{EOM E32}\\
\partial_uE^{3,3}_{ab}&=-\f{3}{20}\big(\Delta+4R\big)E^{2,3}_{ab},\\
\partial_uE^{3,4}_{ab}&=0,
\ee
\esub
where the STF part is defined as $2D_{\la a}\P_{b\ra}=2D_{(a}\P_{b)}-q_{ab}(D_c\P^c)=D_a\P_b+D_b\P_a-q_{ab}(D_c\P^c)$. In the absence of logarithmic terms the evolution equations \eqref{EOM E20} and \eqref{EOM E30} can be found respectively in equations (2.19) and (2.20) of \cite{Geiller:2024bgf}. Here these equations are corrected by the logarithmic contributions $D_{ab}$ and $E^{n,m\geq1}_{ab}$, but we do not write them down explicitly because they are not necessary for our main message.

\subsection{Expansion of the spin coefficients}

The spin coefficients are defined from the tetrad \eqref{null frames} as $\gamma_{ijk}\coloneqq e^\mu_je^\nu_k\nabla_\nu e_{i\mu}$. Some of the coefficients needed in the main text have the definition and expansion
\bsub\label{spin coefficients expansion}
\be
\alpha&=\f{1}{2}(\gamma_{124}-\gamma_{344})=\f{\alpha_1}{r}+\O(r^{-2}),\\
\beta&=\f{1}{2}(\gamma_{123}-\gamma_{343})=-\f{\bar{\alpha}_1}{r}+\O(r^{-2}),\\
\gamma&=\f{1}{2}(\gamma_{122}-\gamma_{342})=-\f{\partial_u\epsilon_2}{r}+\O(r^{-2}),\label{spin gamma}\\
\epsilon&=\f{1}{2}(\gamma_{121}-\gamma_{341})=\f{\epsilon_2}{r^2}+\O(r^{-3}),\\
\mu&=\gamma_{423}=\f{R}{4r}+\O(r^{-2}),\label{spin mu}\\
\nu&=\gamma_{422}=-\f{\ethb R}{4r}+\O(r^{-2}),\label{spin nu}\\
\lambda&=\gamma_{424}=\f{\partial_u\bar{\sigma}_2}{r}+\O(r^{-2}),\\
\tau&=\gamma_{132}=\f{\ethb\sigma_2}{r^2}+\O(r^{-3}),\label{spin tau}\\
\sigma&=\gamma_{133}=\f{\sigma_2}{r^2}+\O(r^{-3}),\label{spin sigma}
\ee
\esub
where we have denoted
\be\label{alpha sigma epsilon}
\alpha_1=\f{1}{2}D_a\bar{m}^a_1,
\q\q
\sigma_2=-\f{1}{2}C_{ab}m^a_1m^b_1,
\q\q
\epsilon_2=\f{1}{4}(\sigma_2-\bar{\sigma}_2).
\ee
Furthermore, we have introduced the spin-weighted derivative acting as
\be
\ethb\F_s\coloneqq\big(\bar{m}^a_1\partial_a+2s\alpha_1\big)\F_s
\ee
on a functional $\F_s$ of spin $s$.

\section{Renormalization}
\label{app:renormalization}

The usual argument for symplectic renormalization via corner terms is to observe that the divergent part of the radial potential can always be written as the sum of a corner term, a total variation, and an angular boundary term \cite{Freidel:2019ohg,McNees:2023tus,Freidel:2024tpl,Geiller:2022vto,Geiller:2024amx,McNees:2024iyu}. This comes from the fact that the on-shell Lagrangian is
\be\label{variation Lagrangian}
\delta L\hateq~\partial_\mu\theta^\mu=\partial_u\theta^u+\partial_r\theta^r+\partial_a\theta^a.
\ee
Here one can indeed see from \eqref{theta components u and r} that
\be\label{theta r div}
\theta^r_\text{div}=-\partial_u\theta^u_0-\delta\big(\sq\,R\big),
\ee
which shows that the $r$-divergent part of the radial potential is itself a corner term up to a total variation. This corner term can then be used to renormalize the $r$-divergent part of the charges. Dropping the irrelevant total variation, we choose this corner term in the form
\be
\vartheta_\text{ren}\coloneqq-\theta^u_0=\f{1}{2}\sq\,q^{ab}\delta C_{ab}.
\ee
To see that this corner potential does indeed renormalize the charges \eqref{bare charge}, we compute
\be\label{Qdiv from corner term}
\xi\ipp(\delta\vartheta_\text{ren})=\f{1}{2}\delta_\xi\big(\sq\,q^{ab}\big)\delta C_{ab}-\f{1}{2}\delta\big(\sq\,q^{ab}\big)\delta_\xi C_{ab}=\slashed{\delta}Q_\text{div}+\sq\,D_a\Upsilon^a,
\ee
where the boundary term is
\be
\Upsilon^a
&=Y^aC^{bc}\delta q_{bc}-q^{ab}\delta C_{bc}Y^c-\f{1}{2}\delta q^{ab}C_{bc}Y^c-\f{1}{2}C^{ab}\delta q_{bc}Y^c\cr
&\pe-fD_b\delta q^{ab}+\delta q^{ab}\partial_bf-2f\partial^a\delta\ln\sq+2\partial^af\delta\ln\sq.
\ee
In order to derive this result we have used the identities
\bsub
\be
D_aD_b\delta q^{ab}&=-\delta R-R\delta\ln\sq-2\Delta\delta\ln\sq,\\
\sq\,\delta q^{ab}D_{\la a}\partial_{b\ra}f
&=-f\delta(\sq\,R)-\Delta f\delta\sq\cr
&\pe+\sq\,D_a\big(\delta q^{ab}\partial_bf-D_b\delta q^{ab}f-2fD^a\delta\ln\sq+2\partial^af\delta\ln\sq\big).
\ee
\esub
Up to the irrelevant total derivative, \eqref{Qdiv from corner term} does indeed show that the divergent part \eqref{Q div} is controlled by the corner term $\vartheta_\text{ren}$.

The radial part of the potential $\theta^r$ has no logarithmically-divergent contribution. However, in order to renormalize the logarithmic contribution to the charge we can use the $\theta_1^u$ part of the temporal potential. Indeed, using the corner term
\be
\vartheta_\text{ln-ren}\coloneqq-\theta^u_1=\sq\,q^{ab}\delta D_{ab},
\ee
one can check that
\be
\xi\ipp(\delta\vartheta_\text{ln-ren})=\delta_\xi\big(\sq\,q^{ab}\big)\delta D_{ab}-\delta\big(\sq\,q^{ab}\big)\delta_\xi D_{ab}=\slashed{\delta}Q_\text{ln-div}+\sq\,D_a\Upsilon^a,
\ee
where the boundary term is
\be
\Upsilon^a=2Y^aD^{bc}\delta q_{bc}-2q^{ab}\delta D_{bc}Y^c-\delta q^{ab}D_{bc}Y^c-D^{ab}\delta q_{bc}Y^c.
\ee
This example illustrates the generality of symplectic renormalization even in the case where the radial potential $\theta^r$ does not exhibit the same divergent structure as the bare charges.

\section{Christodoulou's argument}
\label{Christo argument}

In this appendix we briefly summarize Christodoulou's argument relating the $1/u$ mode of the shear to the loss of peeling \cite{2002nmgm.meet...44C}. For this, we first translate between our notations and conventions (which are commonly used in recent literature on asymptotic symmetries and soft theorems) and those used in \cite{2002nmgm.meet...44C} (which are inherited from the \textit{magnum opus} \cite{Christodoulou:1993uv}). Part of this dictionary between Christodoulou--Klainerman and Newman--Penrose can also be found in appendix A.3 of \cite{Kehrberger:2024aak}.

Let us first compute the shear of the outgoing and ingoing null vectors $\ell$ and $n$ defined in \eqref{null vectors}. For this we compute the extrinsic curvatures
\be\label{curvatures}
\!\!\!K_{ab}(\ell)=\f{1}{2}\pounds_\ell g_{ab}=\f{1}{2}\partial_r\gamma_{ab},
\quad
K_{ab}(n)=\f{1}{2}\pounds_ng_{ab}=\f{1}{2}e^{-2\beta}\left(\partial_u\gamma_{ab}+2\gamma_{(ac}\D_{b)}U^c+\f{V}{2r}\partial_r\gamma_{ab}\right).
\ee
These are called respectively $\chi$ and $\underline{\chi}$ in \cite{2002nmgm.meet...44C}. Taking the trace-free parts in the transverse metric $r^{-2}\gamma_{ab}$, we then obtain the shears
\be\label{shears}
S_{ab}(\ell)=-\f{1}{2r^2}C_{ab}+\O(r^{-3}),
\q\q
S_{ab}(n)=\f{1}{2r}N_{ab}+\O(r^{-2}),
\ee
which in \cite{2002nmgm.meet...44C} are respectively called $\hat{\chi}$ and $\hat{\underline{\chi}}$. The rest of the dictionary between the notations is gather in the following table.
\begin{longtable}{|c|c|c|} 
\hline
& Christodoulou \cite{2002nmgm.meet...44C} & $\q$ here $\q$ \\ \hline
outgoing null normal & $L$ & $\displaystyle\Big.\ell$ \eqref{null vectors} \\ \hline
ingoing null normal & $\underline{L}$ & $\displaystyle\Big.n$ \eqref{null vectors} \\ \hline
normalisation & $L\cdot\underline{L}=-2$ & $\displaystyle\Big.\ell\cdot n=-1$ \\ \hline
fundamental form & $\chi$ & $\displaystyle\Big.K(\ell)$ \eqref{curvatures} \\ \hline
fundamental form & $\underline{\chi}$ & $\displaystyle\Big.K(n)$ \eqref{curvatures} \\ \hline
trace-free part & $\displaystyle\Bigg.\hat{\chi}=\f{\Sigma}{r^2}+\O(r^{-3})$ & $S(\ell)$ \eqref{shears} \\ \hline
trace-free part & $\displaystyle\Bigg.\hat{\underline{\chi}}=\f{\Xi}{r}+\O(r^{-2})$ & $S(n)$ \eqref{shears} \\ \hline
shear & $\Sigma$ & $\displaystyle\Bigg.C_{ab}\quad\to\quad\sigma_2=-\f{1}{2}C_{ab}m^a_1m^b_1$ \\ \hline
news & $\Xi=-2\partial_u\Sigma$ & $\displaystyle\bigg.N_{ab}=\partial_uC_{ab}\quad\to\quad\bar{\lambda}_1=\partial_u\sigma_2$ \\ \hline
Weyl scalar & $\displaystyle\Bigg.\underline{\alpha}=\f{\underline{A}}{r}+\O(r^{-2})$ & $\displaystyle\Psi_4=\f{\Psi_4^0}{r}+\O(r^{-2})$ \eqref{Psi4} \\
& $\underline{A}=-2\partial_u\Xi$ & $\Psi_4^0=\N_{ab}\bar{m}^a\bar{m}^b=-\partial_u\lambda_1$ \eqref{curly N} \\ \hline
Weyl scalar & $\displaystyle\Bigg.\underline{\beta}=\f{\underline{B}}{r^2}+\O(r^{-3})$ & $\displaystyle\Psi_3=\f{\Psi_3^0}{r^2}+\O(r^{-3})$ \eqref{Psi3} \\
& $\underline{B}={}^\circ\slashed{\text{div}}~\Xi$ & $\displaystyle\Bigg.\Psi_3^0=\J_a\bar{m}^a=-\eth\lambda_1+\f{1}{4}\ethb R$ \eqref{curly J} \\ \hline
Weyl scalar & $\displaystyle\Bigg.\rho+i\sigma=\f{P+iQ}{r^3}+\O(r^{-4})$ & $\displaystyle\Psi_2=\f{\Psi_2^0}{r^3}+\O(r^{-4})$ \eqref{Psi2} \\
& & $\displaystyle\Big.\Psi_2^0=\M+i\widetilde{\M}$ \eqref{curly M}\eqref{curly M dual} \\ \hline
Weyl scalar & $\displaystyle\Bigg.\beta=\f{B_*}{r^4}(\ln r)+\f{B}{r^4}+\O(r^{-5})$ & $\displaystyle\Psi_1=\f{\Psi_4^{0,1}}{r^4}(\ln r)+\f{\Psi_4^0}{r^4}+\O(r^{-5})$ \eqref{Psi1} \\
& $B_*=-\,{}^\circ\slashed{\text{div}}~A_*$ & $\Psi_1^{0,1}=D^bD_{ab}m^a\quad\ \,\Psi_1^0=\P_am^a$ \eqref{curly P} \\ \hline
Weyl scalar & $\displaystyle\Bigg.\alpha=\f{A_*}{r^4}+\O(r^{-5})$ & $\displaystyle\Psi_0=\f{\Psi_0^{-1}}{r^4}+\O(r^{-5})$ \eqref{Psi0} \\ \hline
Bianchi identity & equation (3) $\displaystyle\Bigg.\underline{D}\beta=\f{R}{r^4}+\O(r^{-5})$ & \eqref{NP1} \\ \hline
EOM for $\Psi_1^0$ & equation (6) $\displaystyle\Bigg.\partial_uB=\f{1}{2}R$ & $\partial_u\Psi_1^0=\eth\Psi_2^0-2\sigma_2\Psi_3^0\eqqcolon\R$ \eqref{Psi10 EOM} \\ \hline
EOM for $\Big.\Psi_1^{0,1}$ & equation (6) $\partial_uB_*=0$ & $\partial_u\Psi_1^{0,1}=0$ \eqref{Psi10 EOM} \\ \hline
EOM for $\Big.\Psi_2^0$ & equations (10) and (11) & $\partial_u\Psi_2^0=\eth\Psi_3^0-\sigma_2\Psi_4^0$ \eqref{Psi20 EOM} \\ \hline
limits for $\R$ & equation (9) $\displaystyle\Big.R^\pm=\lim_{u\to\pm\infty}uR$ & $\displaystyle\R^\pm=\lim_{u\to\pm\infty}u\R$ \\ \hline
spin coefficient & $\displaystyle\Bigg.\hat{\underline{\chi}}-\f{1}{2}\tr\underline{\chi}+2\underline{\omega}$ & $\mu-\gamma$ \eqref{spin gamma}\eqref{spin mu} \\ \hline
spin coefficient & $\underline{\xi}$ & $\displaystyle\Big.\nu$ \eqref{spin nu} \\ \hline
spin coefficient & $\zeta$ & $\displaystyle\Big.\tau$ \eqref{spin tau} \\ \hline
spin coefficient & $\hat{\chi}$ & $\displaystyle\Big.\sigma$ \eqref{spin sigma} \\ \hline
derivative & $D$ & $\displaystyle\Big.\ell^\mu\partial_\mu$ \\ \hline
derivative & $\underline{D}$ & $\displaystyle\Big.n^\mu\partial_\mu$ \\ \hline
derivative & $\slashed{\nabla}$ & $\displaystyle\Big.m^\mu\partial_\mu$ \\ \hline
\end{longtable}

With this dictionary at hand, we can now go through Christodoulou's key argument relating the $1/u$ mode of the shear (arising from massive particles at $i^-$ which radiate in accordance with the quadrupole formula in post-Newtonian expansion) to the loss of peeling in $\Psi_1$. Once again, we stress that this argument has been analyzed much more meticulously by Kehrberger (see \cite{Kehrberger:2024clh} and section 1.2 of \cite{Kehrberger:2021uvf}). We want nonetheless to reproduce the steps of \cite{2002nmgm.meet...44C} because this construction can also be used to argue why the subleading $(\ln u)/u^2$ terms in \eqref{u expansion of e} are compatible with the $\ln r$ terms appearing in \eqref{Psi0}. Although heuristic, this argument serves to justify the choice of polyhomogeneous expansion \eqref{angular metric} (more precisely the fact that $m_\text{max}=n+1$) from the results of Saha, Sahoo and Sen \cite{Sahoo:2018lxl,Saha:2019tub}. The steps of the argument are as follows:
\begin{enumerate}
\item Consider a fixed round sphere metric so that the Ricci scalar of $q_{ab}$ is $R=2$ (as listed in the table above, this is \textit{not} the $R$ of \cite{2002nmgm.meet...44C}, which we denote here by $\R$ instead) and $\Psi_3^0=-\eth\lambda_1$.
\item Suppose that the shear and the news (by virtue of the definition $\lambda_1=\partial_u\bar{\sigma}_2$) behave as\footnote{Here the subscript 2 on $\sigma_2$ refers to the order of the radial expansion \eqref{spin sigma}, while the superscript 1 on $C$ refers to the order of the $u$ expansion as in \eqref{general C falloffs}.}
\be
\lim_{u\to-\infty}u\sigma_2\eqqcolon C^1\neq0,
\q\q
\lim_{u\to-\infty}u^2\lambda_1=-\bar{C}^1.
\ee
This input, which introduces precisely the $1/u$ tail to the memory in the shear as in \eqref{general C falloffs}, is derived in \cite{2002nmgm.meet...44C} from the quadrupolar approximation applied to massive particles in the far past.
\item Using the Bianchi identity \eqref{Psi20 EOM}, the evolution equation for $\Psi_2^0$ takes the form
\be
\partial_u\Psi_2^0=\eth\Psi_3^0-\sigma_2\Psi_4^0=-\eth^2\lambda_1+\sigma_2\partial_u\lambda_1,
\ee
and can be integrated to obtain
\be
\lim_{u\to-\infty}u\Psi_2^0=-\eth^2\bar{C}^1.
\ee
\item Using the Bianchi identity \eqref{Psi10 EOM}, the evolution equation for $\Psi_1^0$ takes the form
\be
\partial_u\Psi_1^0=\eth\Psi_2^0-2\sigma_2\Psi_3^0\eqqcolon\R,
\ee
with
\be
\lim_{u\to-\infty}u\R=\R^-=-\eth^3\bar{C}^1,
\ee
which in turn implies crucially that $\Psi_1^0\sim\ln u+(\dots)$.
\item Then, we can integrate from $u_1=t_1-r$ to $u=t-r$ with $t_1=0$ to finally obtain
\be
r^4\big(\Psi_1(u)-\Psi_1(u_1)\big)\sim-\int_{u_1}^u\f{\eth^3\bar{C}^1}{u'}\,\de u'\sim-\eth^3\bar{C}^1\big(\ln|u|-\ln r\big).
\ee
This does indeed imply the loss of peeling since $\Psi_1=\Psi_1^{0,1}(\ln r)r^{-4}+\O(r^{-4})$ instead of the peeling behavior \eqref{Sachs peeling}.
\end{enumerate}
In the main text, working with the transverse metric \eqref{angular metric} has enabled us to obtain the explicit expansion \eqref{Psi1} which relates the symmetric trace-free tensor $D_{ab}$ to $\bar{C}^1$. We have also emphasized that the polyhomogeneous expansion \eqref{angular metric} has been chosen with $m_\text{max}=n+1$. This was done for the sake of generality, in order to illustrate the structure of the logarithmic flux-balance laws, and most importantly in order to accommodate the results of \cite{Kehrberger:2024aak} which illustrate the appearance of $E_{ab}^{1,2}$. We can now also argue along the lines of \cite{2002nmgm.meet...44C} why the presence of such a term is compatible with the subleading universal logarithmic contributions in \eqref{u expansion of e} \cite{Sahoo:2018lxl,Saha:2019tub}.

Consider once again the steps described above, but now with the inclusion of the term $(\ln u)/u^2$ in the shear in accordance with Sahoo and Sen's result \eqref{u expansion of e} \cite{Sahoo:2021ctw}. Let us therefore write
\be
\sigma_2=\f{1}{u}C^1+\f{1}{u^2}\big(C^{2,0}+C^{2,1}(\ln u)\big)+\O\big(u^{-3}\big).
\ee
We can then integrate the expansion of the Bianchi identity 
\eqref{expNP0} to find that 
\be
\Psi_0^0=u\eth^4\bar{C}^1+(\ln u)\eth^4\big(\bar{C}^{2,0}+\bar{C}^{2,1}-u\bar{C}^1\big)+\f{1}{2}(\ln u)^2\eth^4\bar{C}^{2,1}+\O\big(u^{-1}\big).
\ee
To close the heuristic argument, we can appeal to the same reasoning as above to connect the appearance of $(\ln u)$ to $(\ln r)$. This is because the logarithmic terms actually involve $\ln(u/u_1)$, and we are taking $u_1=t_1-r$ with $t_1=0$. If we now substitute
\be\label{Christo subleading}
(\ln u)\eth^4\big(\bar{C}^{2,0}+\bar{C}^{2,1}-u\bar{C}^1\big)+\f{1}{2}(\ln u)^2\eth^4\bar{C}^{2,1}\quad\to\quad(\ln r)\eth^4\big(\bar{C}^{2,0}+\bar{C}^{2,1}-u\bar{C}^1\big)+\f{1}{2}(\ln r)^2\eth^4\bar{C}^{2,1},
\ee
we can compare the right-hand side with \eqref{Psi0}, and note that the logarithmic terms have the same structure. More precisely, we know from the evolution equations \eqref{EOM D and Enn1} and \eqref{EOM E11 and Enn} that $E^{1,2}_{ab}$ is constant in $u$ while $E^{1,1}_{ab}$ is linear in $u$, but this corresponds precisely to the result of the argument \eqref{Christo subleading}. In summary, we have here extended the argument of Christodoulou to connect not only the $1/u$ tail in the shear to the loss of peeling in $\Psi_1$, but also the subleading structure of the shear to the subleading logarithmic terms in $\Psi_0$. This is an argument in favor of the ansatz \eqref{angular metric} with $m_\text{max}=n+1$ in order to describe the generic scattering situations as in \cite{Sahoo:2018lxl,Saha:2019tub}.

\bibliography{Biblio.bib}
\bibliographystyle{Biblio}

\end{document}